\documentclass{ijuc}
\usepackage[english]{babel}
\usepackage{graphicx}

\newcommand {\ket} [1] {| #1 \rangle}
\newcommand {\bra} [1] {\langle #1 |}

\newcommand {\mh}  {\hspace{.3 in}}
\newcommand {\sh}  {\hspace{.1 in}}
\newcommand {\sv}  {\smallskip}

\newcommand {\qq} {``}

\begin{document}



\title{Evolution of Quantum Systems\\by Diagrams of States}


\author{Sara Felloni$^{1,2}$\email{sara.felloni@iet.ntnu.no}
     \and Alberto Leporati$^3$
     \and Giuliano Strini$^4$
  }

\institute{The Norwegian University of Science and Technology (NTNU)\\
Department of Electronics and Telecommunications\\
NO-7491, Trondheim, Norway\\
\sv
\and
UNIK - University Graduate Center,
NO-2027 Kjeller, Norway
\sv
\and
Dipartimento di Informatica, Sistemistica e Comunicazione \\
   Universit\`a degli Studi di Milano -- Bicocca\\
Viale Sarca 336/14, 20126 Milano, Italy\\
\sv
\and
Dipartimento di Fisica\\
Universit\`a degli Studi di Milano\\
Via Celoria 16, 20133 Milano, Italy\\
}

\maketitle

\begin{abstract}


We explore the main processes involved in the evolution of general quantum systems by means of \textit{Diagrams of States}, a novel method to graphically represent and analyze how quantum information is elaborated during computations performed by quantum circuits.
We present quantum diagrams of states for representations of quantum states by density matrices, partial trace operations, density matrix purification and time-evolution by Kraus operators. Following these representations, we describe by diagrams of states the most general transformations related to single-qubit decoherence and errors.\\
Diagrams of states prove to be a useful approach to analyze quantum computations, by offering an intuitive graphic representation of the processing of quantum information. They also help in conceiving novel quantum computations, from describing the desired information processing to deriving the final implementation by quantum gate arrays.

\end{abstract}


\section{Evolution of Quantum Systems by Graphic Representation of States}

We explore the main processes involved in the evolution of general quantum systems by means of \textit{Diagrams of States}. The representation by diagrams of states is a novel method to graphically represent and analyze how quantum information is elaborated during computations performed by quantum circuits. In the widely-used representation by quantum circuits, horizontal lines represent single qubits constituting the considered quantum system. In contrast, in diagrams of states we draw a horizontal line for each state of the computational basis. Therefore, diagrams of states are less synthetic in respect to quantum circuits, but allow a clear and straightforward visualization of the quantum information processing.

We previously introduced this method by defining basic representations for standard quantum operations and providing examples of basic practical quantum computations \cite{FeStsdI08}. In this paper, diagrams of states are applied to explore the density matrix representation of quantum states, the time-evolution of open quantum systems by Kraus operators, and a general model for single-qubit decoherence and errors. As in previous related works \cite{FeStsdI08, FeSt06}, diagrams of states will be used both as a novel approach to investigate quantum computations, in addition to (or in substitution of) standard methods like analytical study and Feynman diagrams \cite{Fey88, Fey99}, and as an auxiliary tool to construct novel quantum computations from the desired manipulation of quantum states.

This paper is organized as follows. In Section~\ref{sec-sd-II-denskraus}, we consider the representation of quantum states by density matrices, and their partial trace and purification processes for composite systems; as practical examples, we illustrate physically feasible procedures to reduce density matrices in two-qubit systems and to purify density matrices of single-qubit systems. We conclude this section with the Kraus representation for the most general time-evolution of density matrices. In Section~\ref{sec-sd-II-deco1q}, we explore the most general transformation to model single-qubit decoherence and errors \cite{BeFeSt06}.
Finally, in Section~\ref{sec-sd-II-concl} we present our conclusions.

Throughout this paper, in order to perform the analysis of given quantum processes, we shall directly derive diagrams of states from the quantum circuits associated with the physical implementation of the processes. These diagrams can easily be rearranged into new simpler diagrams, which better visualize the overall manipulation of information from input to output: We shall refer to the former as \textit{complete} diagrams and to the latter as \textit{simplified} diagrams.

Any sequence of logic gates must be read from left (input) to right (output), both for conventional quantum circuits and for their representations by means of diagrams of states. From top to bottom, qubits run from the least significant (\textsc{lsb}) to the most significant (\textsc{msb}).

\section{Density Matrices and the Kraus Representation}\label{sec-sd-II-denskraus}

The density matrix representation (see, \textit{e.g.}, \cite{qcbook2}, pages 258-264, or \cite{CoDiLa77}, pages 295-307) is very useful whenever representing non-pure states or describing their time evolution. Any quantum state can be represented by means of a density matrix which, in turn, can be expressed by its spectral decomposition:
\begin{equation}
    \rho = \sum_i \lambda_i \ket i \bra i,
\end{equation}
where $\{\ket i\}$ are the density matrix eigenvectors and $\{\lambda_i\}$ are the corresponding eigenvalues.

\subsection{Partial trace and reduced density matrices}

The Hilbert space of a composite system which is constituted by two independent or interacting subsystems can be obtained as the tensor product of the Hilbert spaces corresponding to the two constitutive subsystems.

Given the density matrix of a general state of the composite system, a partial trace operation consists in finding a description of states which considers only elements related to a certain subsystem (denoted as subsystem $I$) while disregarding any element related to the remaining subsystem (denoted by $II$):
\begin{equation}
    \rho^I \equiv Tr_{II} \{\rho\} \equiv \sum_i \bra i \rho \ket i,
\end{equation}
where $i$ is the index corresponding to subsystem $II$.

\subsubsection{Reduced density matrices in the spaces of two qubits}

For the sake of simplicity, we consider a system composed of two single-qubit subsystems, whose general overall state is described by a density matrix of dimension $4 \times 4$. By performing on this density matrix the partial trace operations on each one of the two single-qubit subsystems, it is possible to define the two corresponding reduced density matrices (of dimension $ 2 \times 2 $).

\begin{figure}[!htb]
\begin{center}
\includegraphics[width=9.4cm]{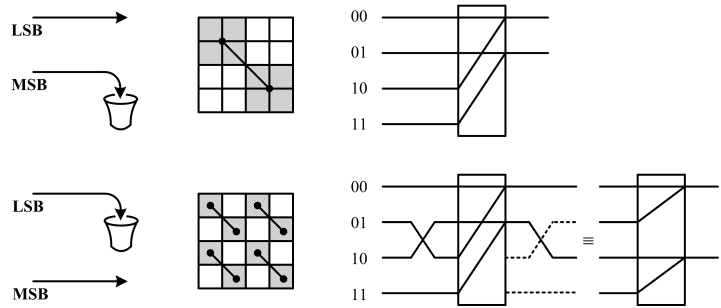}
\end{center}
\caption{Graphic representations of partial trace operations for a two-qubit system, on the most significant qubit (upper figures) and on the least significant qubit (lower figures). From left to right: quantum circuits, additions of matrix entries, diagrams of states. The reduced density matrix of the subsystem constituted by the least significant qubit is obtained by adding the $ 2 \times 2 $ sub-matrices highlighted in gray shading. Similarly, the reduced density matrix of the subsystem constituted by the most significant qubit is obtained by adding in pairs the entries highlighted in gray shading.}
\label{sd-II-partialtr}
\end{figure}

The two possible partial trace operations are graphically represented in Figure \ref{sd-II-partialtr}. The processing of information involved in the partial trace operations is most clearly illustrated in the diagrams of states (left). Evidently, the diagram related to the partial trace on the least significant qubit (lower) can be directly derived from the diagram related to the partial trace on the most significant qubit (upper) by appropriately permuting the qubits constituting the system; the permutation is obtained by simply applying swap gates before and after the partial trace operation.
By appropriate immersions and permutations of diagrams of states \cite{FeStsdI08}, the method can be easily generalized to describe partial trace operations in systems of higher dimensionality and in respect to subsystems constituted by any combination of qubits.
Finally, the diagram-of-states representation for partial trace operations will be a fundamental tool for the description of decoherence and evolution of open single-qubit systems, addressed in Section \ref{sec-sd-II-deco1q}.

\subsection{Purification of a density matrix}

The purification of density matrices (see, \textit{e.g.}, \cite{qcbook2}, pages 267-270) can be considered as a sort of inverse process to partial trace operations.
A general non-pure quantum state in a system denoted by $I$ is represented by its density matrix $\rho^I$. This density matrix $\rho^I$ can be seen as a reduced density matrix obtained by partial trace from a larger density matrix $\rho$, describing a pure state in a larger system. The larger system would include the initial system $I$ and a second ancillary quantum system $II$, and it is denoted by $I + II$. The purification of the density matrix $\rho^I$ consists in determining the quantum system $I + II$ which leads to the reduction process.

By using the Latin set of indices $\{ i, j, ...\}$ when referring to the initial quantum system $I$ and the Greek set of indices $\{\alpha, \beta, ...\}$ when referring to the ancillary quantum system $II$, a general pure state in the overall system $I+II$ is described by:
\begin{equation}
    |\Phi\rangle\equiv\sum_{i \; \alpha} \, C_{i\alpha} \, |i\rangle|\alpha\rangle
\end{equation}
and by the density matrix:
\begin{equation}
    \rho\equiv\sum_{i \, \alpha}\sum_{j \; \beta}C_{i\alpha} \,
    C_{j\beta}^* \, |i\rangle|\alpha\rangle\langle\beta|\langle
    j|.
\end{equation}
The assigned density matrix $\rho^I$ is in general not diagonal:
\begin{equation}\label{eq-sd-II-rhoI}
    \rho^I\equiv\sum_{k \; l}|k\rangle\rho^I_{kl}\langle l|.
\end{equation}
By imposing that the density matrix $\rho^I$ is obtained by partial trace of the density matrix $\rho$ of the overall system in respect to the basis $\{|\gamma\rangle\}$ of the subsystem $II$, we obtain:
\begin{equation}
\rho^I\equiv\sum_{k\;l}|k\rangle\rho^I_{kl}\langle l|=\sum_\gamma
\; \left\{ \sum_{i\;j}C_{i\gamma} \, C_{j\gamma}^* \, |i\rangle\langle
j|\right\},
\end{equation}
that is, a set of equations for the entries of the density matrix $\rho^I$:
\begin{equation}\label{eq-sd-II-coefrhoI}
    \rho^I_{ij}=\sum_\gamma C_{i\gamma}C_{j\gamma}^*,
\end{equation}
where the coefficients $\rho_{ij}^I$ are known, while the coefficients $C_{i\gamma}$ are to be determined.

As is well known, equations (\ref{eq-sd-II-coefrhoI}) are solvable if the space of the basis $\{\ket\gamma\}$ has a sufficiently high dimensionality. That is, given any density matrix in a defined quantum system, it is always possible to consider an extended quantum system in which the assigned non-pure state can be obtained by partial trace of a pure state in the extended system.

\subsubsection{Density matrix purification for a single-qubit system}

For the sake of simplicity, we consider the purification process for a density matrix $\rho$ assigned in a single-qubit system. A single ancillary qubit is sufficient to obtain the corresponding pure state $\ket\Psi$ in a larger quantum system.

From equations (\ref{eq-sd-II-coefrhoI}) we obtain a three-equation system in four unknown $C_{i\alpha}$:
\begin{equation}\label{eq-sd-II-sisrhoI}
\left\{%
\begin{array}{c}
\rho_{11}=C_{00}C_{00}^*+C_{01}C_{01}^*\nonumber\\
\rho_{12}=C_{00}C_{10}^*+C_{01}C_{11}^*=\rho^*_{21}\\
\rho_{22}=C_{10}C_{10}^*+C_{11}C_{11}^*\nonumber\\
\end{array}%
\right.
\end{equation}
We can assume the parameter $C_{00}$ to be real, since a common phase is arbitrary, and we can impose the condition $C_{01}=0$, thus obtaining:
\begin{equation}
  C_{00}=\sqrt\rho_{11}\, , \sh
  C_{01}=0\, , \sh
  C_{10}=\frac{\rho^*_{12}}{\sqrt\rho_{11}}\, , \sh
  C_{11}=\sqrt\frac{\rho_{11}\,\rho_{22}-|\rho_{12}|^2}{\rho_{11}}\, .
\end{equation}
Thus, the following pure two-qubit state performs a purification of the given density matrix $\rho$:
\begin{equation}\label{eq-sd-II-purrho}
\ket\Psi= \sqrt\rho_{11}\;\ket0\ket0
+\frac{\rho^*_{12}}{\sqrt\rho_{11}}\;\ket1\ket0+
\sqrt\frac{\rho_{11}\,\rho_{22}-|\rho_{12}|^2}{\rho_{11}}\;\ket1\ket1.
\end{equation}

\begin{figure}[!htb]
\begin{center}
\includegraphics[width=6.4cm]{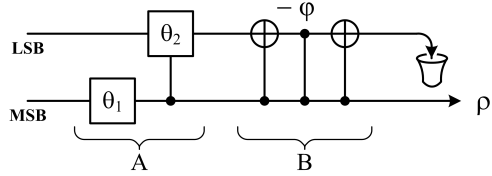}
\end{center}
\caption{Quantum circuit representing the synthesis of a two-qubit state purifying the density matrix of a single-qubit system, by unitary operations performed on a two-qubit register. The gate array \qq A'' synthesizes the amplitude modules, while the gate array \qq B'' synthesizes the phase.}\label{sd-II-cirpurrho}
\end{figure}

\begin{figure}[!htb]
\begin{center}
\includegraphics[width=7.6cm]{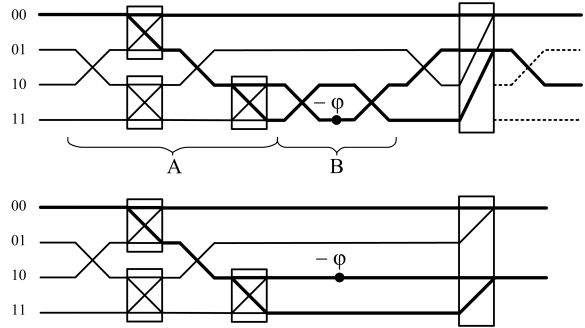}
\end{center}
\caption{Complete (upper) and simplified (lower) diagrams of states representing the purification of the density matrix of a single-qubit system. Starting from the input state $\ket{00}$, the active information is processed along thick lines, while thin lines correspond to absence of information.}\label{sd-II-diastpurrho}
\end{figure}

Figure \ref{sd-II-cirpurrho} shows the quantum circuit to generate the quantum state for the purification of the density matrix $\rho$. The processing of information is clearly illustrated by the complete and simplified diagrams of states in figure \ref{sd-II-diastpurrho}. In these diagrams, the active information contained in the initial states is manipulated from left to right along thick lines, while thin lines correspond to absence of information.

After the application of the \qq A'' gate array of Figures \ref{sd-II-cirpurrho} and \ref{sd-II-diastpurrho}, we obtain the state:
\begin{equation}
\ket{\Psi} = \left[%
\begin{array}{c}
  \cos \theta_1 \\
  0 \\
  \cos \theta_2 \, \sin \theta_1 \\
  \sin \theta_2 \,  \sin \theta_1 \\
\end{array}%
\right].
\end{equation}
Purification is then achieved by applying the \qq B'' gate array, where the controlled -phase gate has phase $\varphi = \arg \,(\rho_{12})$. Thus, the pure state $\ket\Psi$ of equation (\ref{eq-sd-II-purrho}) is generated by two-qubit unitary operations.

\subsection{The Kraus representation}

\begin{figure}[!htb]
\begin{center}
\includegraphics[width=6cm]{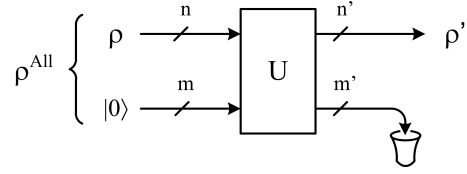}
\end{center}
\caption{Kraus representation for the time-evolution of a general density matrix. The overall number of qubits is preserved during the evolution process represented by the unitary matrix $U$: $n^\prime + m^\prime = n + m$.}\label{sd-II-evolkraus}
\end{figure}

The most general time-evolution of a density matrix can be described by the Kraus representation (\cite{Kra83} or see, \textit{e.g.}, \cite{qcbook2}, pages 278-284).
This representation is schematically illustrated in Figure \ref{sd-II-evolkraus}.

For the sake of simplicity, we consider an evolving quantum system constituted by a single qubit of density matrix $\rho$, and by a single ancillary qubit on which the partial trace operation is performed. This choice causes no loss of generality, as the Kraus representation can easily be generalized for main and auxiliary systems of higher dimensionality by appropriate immersions and permutations of diagrams of states \cite{FeStsdI08}.

\begin{figure}[!htb]
\begin{center}
\includegraphics[width=8.4cm]{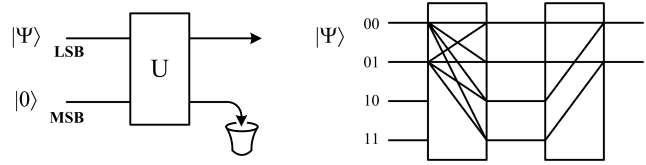}
\end{center}
\caption{Quantum circuit (left) and diagram of states (right) representing the Kraus time-evolution of a single-qubit main quantum system, supported by a single-qubit ancillary system in the most significant position. In the diagram, the active lines of the unitary matrix $U$ determine the evolution of the overall system. A similar representation can be obtained for the ancillary qubit in the least significant position, by simply permuting the qubits constituting the system.}\label{sd-II-diastevkr}
\end{figure}

The processing of information involved in the time-evolution is illustrated by the diagram of states in Figure \ref{sd-II-diastevkr}.
According to Figure \ref{sd-II-evolkraus}, the overall number of qubits is preserved during the evolution process, and in the simplest case considered here we have $n = m = n^\prime = m^\prime = 1$.
The initial density matrix $\rho^{All}$ is given by:
\begin{equation}
    \rho^{All}=\ket0\langle0|\otimes\rho.
\end{equation}
The evolution of the overall density matrix $\rho^{All}$ is then obtained by applying the unitary matrix $U$ appropriately divided into sub-matrices of dimension $2 \times 2$ and, subsequently, by partial tracing on the most significant qubit.
The final density matrix $\rho^\prime$ is:
$$
\rho^\prime=Tr_{MSB}\{U\,\rho^{All}\,U^\dag\}=
$$
$$= Tr_{MSB} \left\{
\left[\begin{array}{cc}A&B\\C&D\end{array}\right]\,
\rho^{All}\,
\left[\begin{array}{cc}A^\dag&C^\dag\\B^\dag&D^\dag\end{array}\right] \right\}=
$$
\begin{equation}\label{eq-sd-II-evolkraus1}
= A\,\rho \,A^\dag+C\,\rho \,C^\dag
\end{equation}
with the condition of trace preservation:
\begin{equation}\label{eq-sd-II-evolkraus1tr}
    A^\dag A+C^\dag C=1.
\end{equation}

\section{Single-Qubit Decoherence and Errors}\label{sec-sd-II-deco1q}

In any realistic physical implementation for quantum computers, quantum systems which process information must always be considered open, that is, never perfectly isolated from the environment. This gives rise to the well known phenomenon of decoherence, with which we here denote any quantum-noise process due to the unavoidable coupling of the main system to the environment (see, \textit{e.g.}, \cite{qcbook2}, pages 335-351).
In addition to decoherence, many sources of imprecision or perturbation have to be taken into account in all quantum information processing tasks, since any experimental implementation of quantum states and operations is unavoidably imperfect to some extent.

To our best knowledge, error models in the literature do not offer a complete and realistic description of the most general noise-transformations for single-qubit systems.
In \cite{BeFeSt06} we considered all physically possible single-qubit errors: A general transformation of a single-qubit density matrix is determined by a set of parameters, with which we associated basic transformations, or quantum noise channels. Here we provide a detailed analysis and complete description of this single-qubit error model, representing all noise channels by means of quantum circuits, diagrams of states and geometrical visualizations of their action on pure quantum states.

\subsection{General transformation of a single-qubit density matrix}

It is well known that any quantum state:
\begin{equation}
    \ket \Psi = \alpha \ket 0 + \beta \ket 1 = \left[
                                               \begin{array}{c}
                                                 \alpha \\
                                                 \beta \\
                                               \end{array}
                                             \right], \sh |\alpha^2| + |\beta^2| = 1,
\end{equation}
can be rewritten in spherical coordinates:
\begin{equation}
\ket \Psi = \cos \frac{\theta}{2} \ket 0 + e^{i \phi }\sin \frac{\theta}{2}  \ket 1 = \left[
                                               \begin{array}{c}
                                                 \cos \frac{\theta}{2} \\
                                                 e^{i \phi }\sin \frac{\theta}{2} \\
                                               \end{array}
                                             \right]
\end{equation}
and represented by the Cartesian coordinates of the three-dimensional space embedding the Bloch sphere.

The quantum state $\ket \Psi$ can also be associated to its density matrix $\rho = \ket \Psi \bra \Psi$ which, in turn, can be expressed by means of Pauli operators $\mbox{\mathversion{bold}$\sigma$\mathversion{normal}}$:
\begin{equation}
    \rho \equiv \frac{1}{2} \, [ I + \mbox{\mathversion{bold}$\lambda$\mathversion{normal}} \mbox{\mathversion{bold}$\sigma$\mathversion{normal}}] = \frac{1}{2} \left[
                         \begin{array}{cc}
                           1 + Z & X - i Y \\
                           X + i Y & 1 - Z \\
                         \end{array}
                       \right],
                       \sh \mbox{\mathversion{bold}$\lambda$\mathversion{normal}} = \left[
                         \begin{array}{c}
                           X \\
                           Y \\
                           Z \\
                         \end{array}
                       \right],
\end{equation}
where $\mbox{\mathversion{bold}$\lambda$\mathversion{normal}}$ is the vector of the Bloch sphere coordinates of the state $\ket \Psi$.

According to the Kraus representation, the most general evolution of a
single-qubit density matrix is given by:
\begin{equation}
    \rho^\prime = \sum_{i} F_{i} \rho F_{i}^{\dagger}, \sh
\sum_{i} F_{i}^{\dagger} F_{i} = I,
\end{equation}
which leads to an affine transformation in the Bloch sphere coordinates of the state $\ket \Psi$:
\begin{equation}
    \mbox{\mathversion{bold}$\lambda$\mathversion{normal}}^\prime = M \mbox{\mathversion{bold}$\lambda$\mathversion{normal}} + \mathbf{c},
\end{equation}
where $M$ is a real matrix of dimension $3 \times 3$ and $\mathbf{c}$ is a translation vector of dimension $3$.
Thus, twelve parameters are necessary and sufficient to characterize a generic
quantum-noise operation acting on a single-qubit system.

These twelve parameters can not be chosen arbitrarily, since they are subjected to the mathematical conditions of complete positivity of the quantum-noise transformations, that is, we always need to assure the physical feasibility of the transformations determined by the parameters. This restriction may require a very difficult treatment if we choose to rely only on analytical descriptions. On the contrary, by studying the parameters by means of equivalent quantum circuits, the mathematical conditions for completely positive transformations are inherently and straightforwardly satisfied, and the same method easily allows for extensions to higher-dimensional cases.

In order to develop descriptions by quantum circuits, we thus decompose the matrix $M$ as one diagonal and two orthogonal matrices, $ M = O_1 \, D \, O_2^T ,$
and we associate the basic parameters with transformations of a general quantum state residing in the Bloch sphere. Three parameters correspond to
rotations of the sphere around the axes $x, y, z$; three parameters correspond to displacements of the sphere along the axes $x, y, z$; three parameters correspond to deformations of the sphere into an ellipsoid with $x$, $y$ or $z$ as symmetry axes. The last three parameters correspond to further rotations of the already deformed sphere, in order to obtain an ellipsoid with symmetry axis along arbitrary directions. Thus, we only need to illustrate in detail the first nine transformations, since deformations of the Bloch sphere into an ellipsoid with symmetry axis along an arbitrary direction can be obtained by appropriate compositions of the first nine transformations.

\subsection{Rotations of the Bloch sphere around the coordinate axes}

Rotation errors are the simplest case of single-qubit errors.
They are unitary transformations and can be obtained immediately as special cases of single-qubit unitary gates, whose diagrams of states are described in detail in \cite{FeStsdI08}.
For the reader's convenience, the corresponding transformations of the Bloch sphere coordinates are:
$$
R_{x} = \left[
\begin{array}{cc}
 \cos \frac{\theta}{2} & - i \sin \frac{\theta}{2} \\
 - i \sin \frac{\theta}{2} & \cos \frac{\theta}{2}
\end{array}\right]
\mh
\left\{
\begin{array}{l}
X^\prime=X\\
Y^\prime=\cos\theta \, Y - \sin \theta \, Z\\
Z^\prime=\sin\theta \, Y + \cos \theta \, Z
\end{array}
\right.
$$
$$
R_{y} = \left[
\begin{array}{cc}
 \cos \frac{\theta}{2} & - \sin \frac{\theta}{2} \\
  \sin \frac{\theta}{2}& \cos \frac{\theta}{2}
\end{array}\right]
\mh
\left\{
\begin{array}{l}
X^\prime=\cos\theta \, X - \sin \theta \, Z\\
Y^\prime=Y\\
Z^\prime=\sin\theta \, X + \cos \theta \, Z
\end{array}
\right.
$$
\begin{equation}
R_{z} = \left[
\begin{array}{cc}
 \cos \frac{\theta}{2} - i \sin \frac{\theta}{2} & 0 \\
0 & \cos \frac{\theta}{2} + i \sin \frac{\theta}{2}
\end{array}\right]
\sh
\left\{
\begin{array}{l}
X^\prime=\cos\theta \, X - \sin \theta \, Y\\
Y^\prime=\sin\theta \, X + \cos \theta \, Y\\
Z^\prime=Z
\end{array}
\right.
\end{equation}
where the error parameter $\theta$ denotes the rotation angle, that is, the intensity of the rotation error applied to the original single-qubit quantum state.

\subsection{Deformations of the Bloch sphere along the coordinate axes}

Deformation errors are non-unitary errors: Consequently, they require a physically feasible representation involving a unitary operation in an extended quantum system.

\begin{figure}[!htb]
\begin{center}
\includegraphics[width=9.4cm]{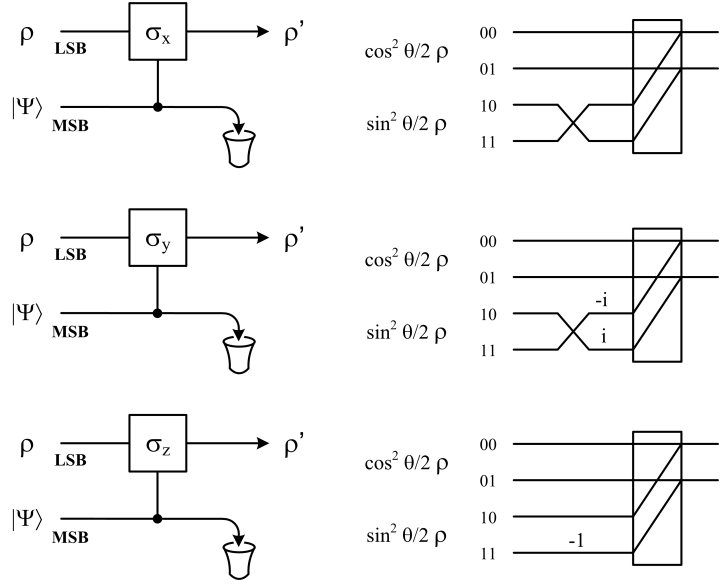}
\end{center}
\caption{Quantum circuits (left) and the corresponding diagrams of states (right) representing deformations of the Bloch sphere along the $x, y$ and $z$ axes. From top to bottom, the bit flip, bit-phase flip and phase flip channels. The ancillary qubit representing the environment is set to the initial state
$|\Psi\rangle= \cos\frac{\theta}{2}|0\rangle+\sin\frac{\theta}{2}|1\rangle$, with
$0\le\theta\le \pi$.}\label{sd-II-deformch}
\end{figure}

The deformation of the Bloch sphere into an ellipsoid centered at the origin of the sphere with axis directed along $x, y$ or $z$ can be respectively implemented by means of the bit flip, bit-phase flip and phase flip channels. The corresponding quantum circuits are illustrated in Figure \ref{sd-II-deformch} (left), where the environment is represented by a single ancillary qubit. The processing of information caused by each deformation channel is clearly illustrated in the corresponding diagrams of states in Figure \ref{sd-II-deformch} (right).

The Kraus operators of each channel can be easily derived from the corresponding diagram of states.
For the bit flip channel we have:
\begin{equation}
F_{1} = \left| \cos\frac{\theta}{2}\right| \, I \mh F_{2} = \left|\sin\frac{\theta}{2}\right| \, \sigma_{x}.
\end{equation}
For the bit-phase flip channel we have:
\begin{equation}
F_{1} = \left|\cos\frac{\theta}{2}\right| \, I \mh F_{2} = \left|\sin\frac{\theta}{2}\right| \, \sigma_{y}.
\end{equation}
Finally, for the phase flip channel we have:
\begin{equation}
F_{1} = \left|\cos\frac{\theta}{2}\right| I \mh F_{2} = \left|\sin\frac{\theta}{2}\right| \, \sigma_{z}.
\end{equation}

The Kraus operators of the three channels induce the following transformations in the Bloch sphere coordinates of the initial state:
\begin{equation}
\left\{
\begin{array}{l}
X^\prime=X\\
Y^\prime=\cos \theta \, Y\\
Z^\prime=\cos \theta \, Z
\end{array}
\right.
\mh
\left\{
\begin{array}{l}
X^\prime=\cos \theta \, X\\
Y^\prime=Y\\
Z^\prime=\cos \theta \, Z
\end{array}
\right.
\mh
\left\{
\begin{array}{l}
X^\prime=\cos \theta \, X\\
Y^\prime=\cos \theta \, Y\\
Z^\prime=Z
\end{array}
\right.
\end{equation}

\begin{figure}[!htb]
\begin{center}
\includegraphics[width=2.4cm]{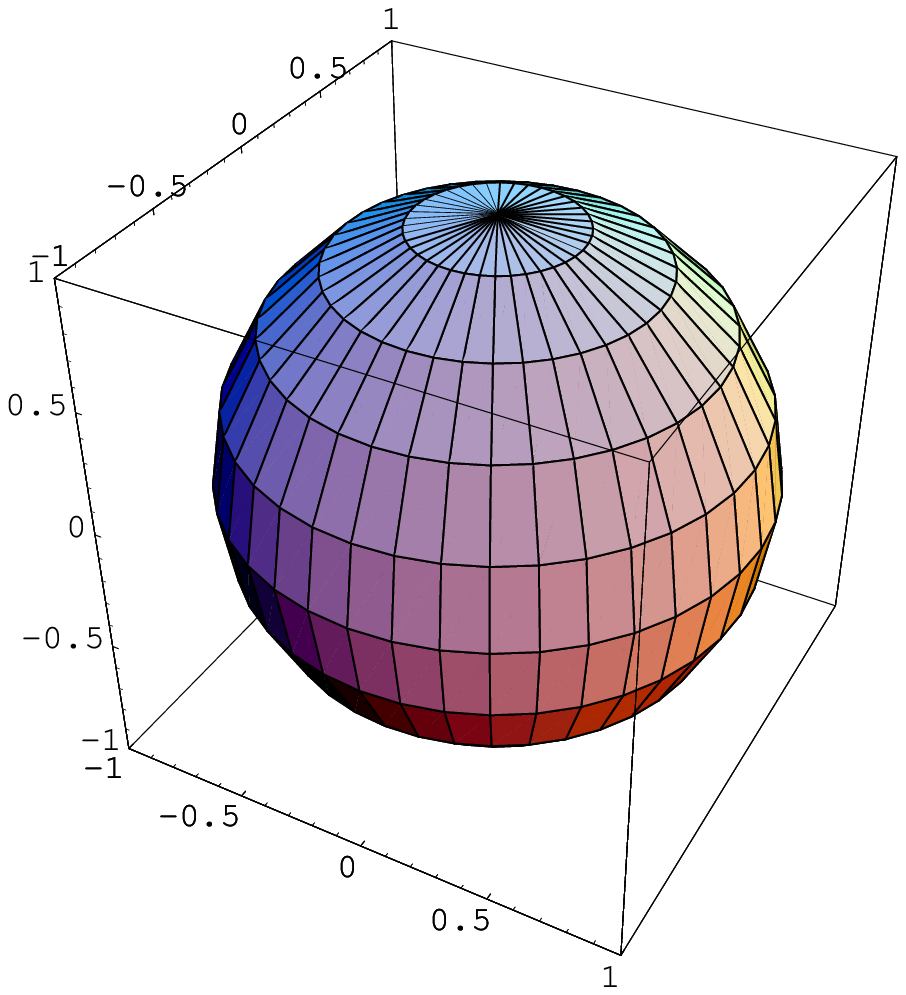}
\includegraphics[width=2.4cm]{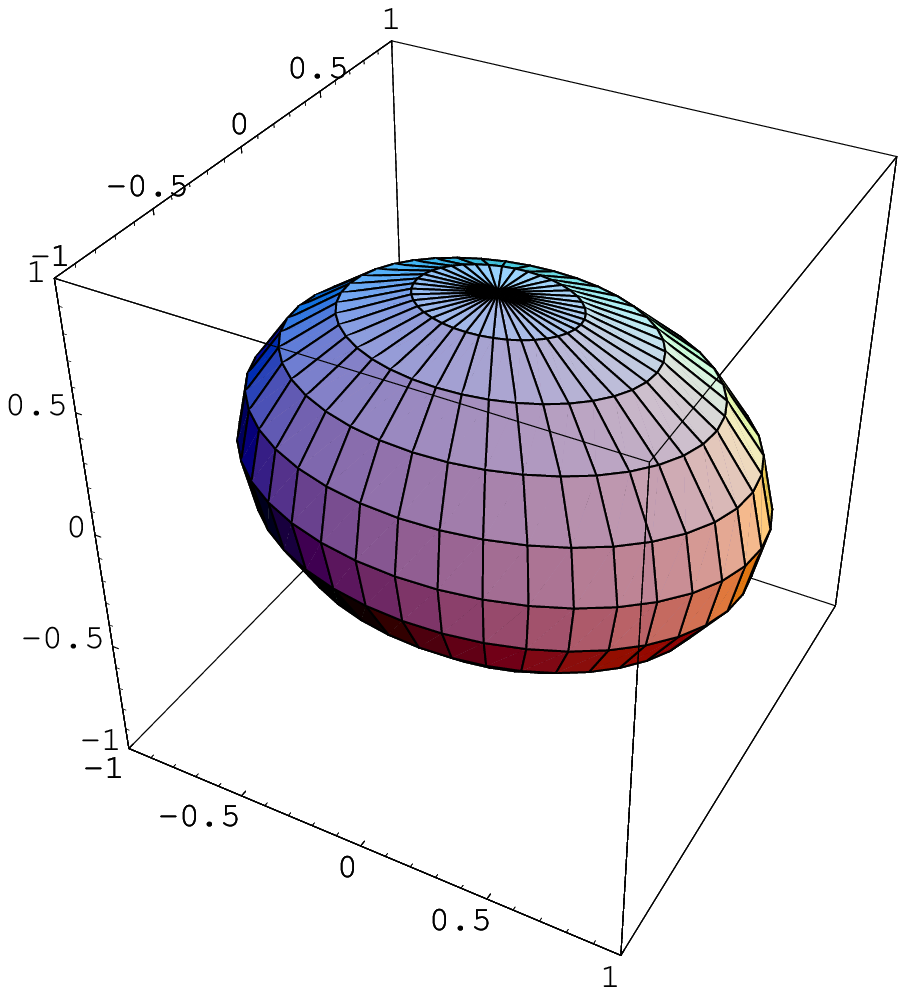}
\includegraphics[width=2.4cm]{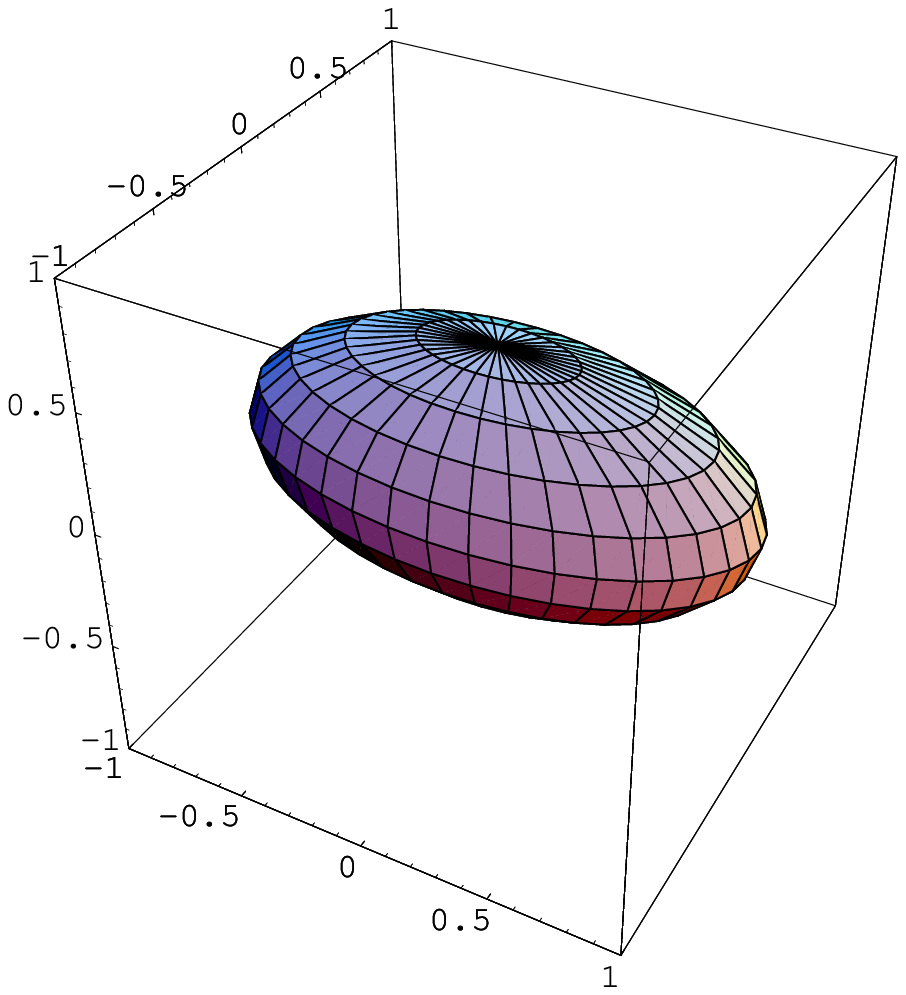}
\\
\includegraphics[width=2.4cm]{SferaBloch.eps}
\includegraphics[width=2.4cm]{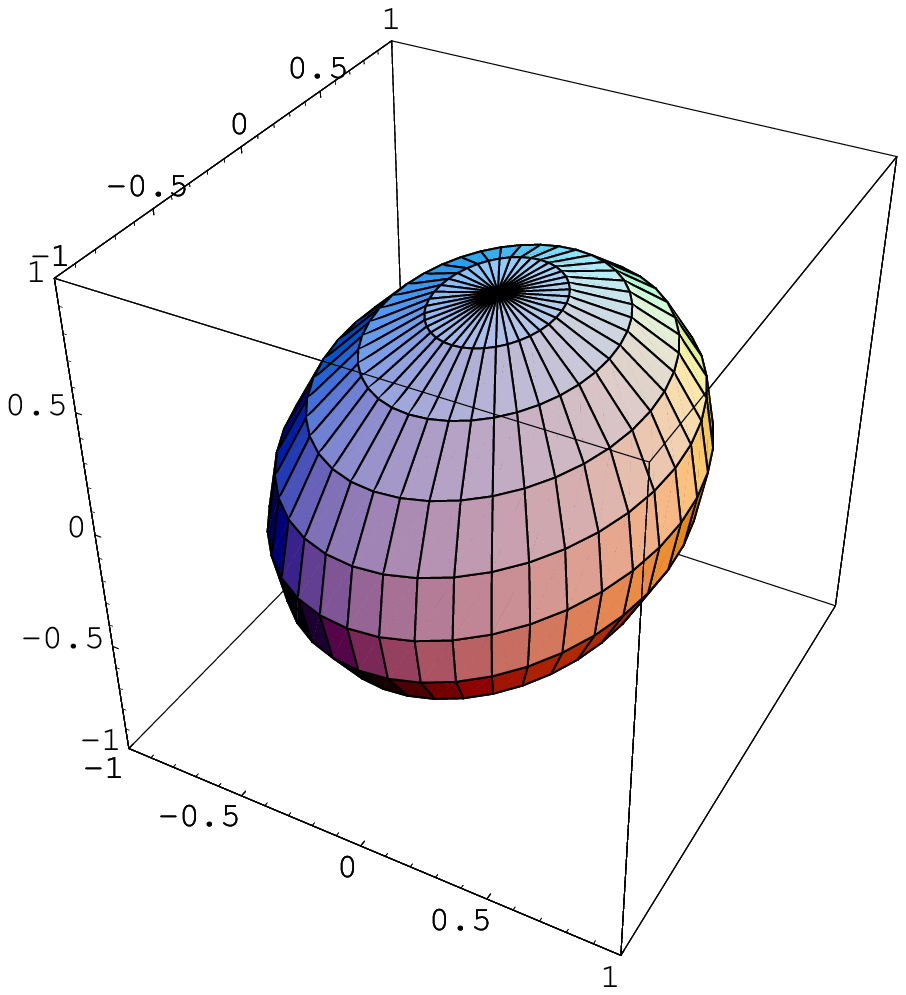}
\includegraphics[width=2.4cm]{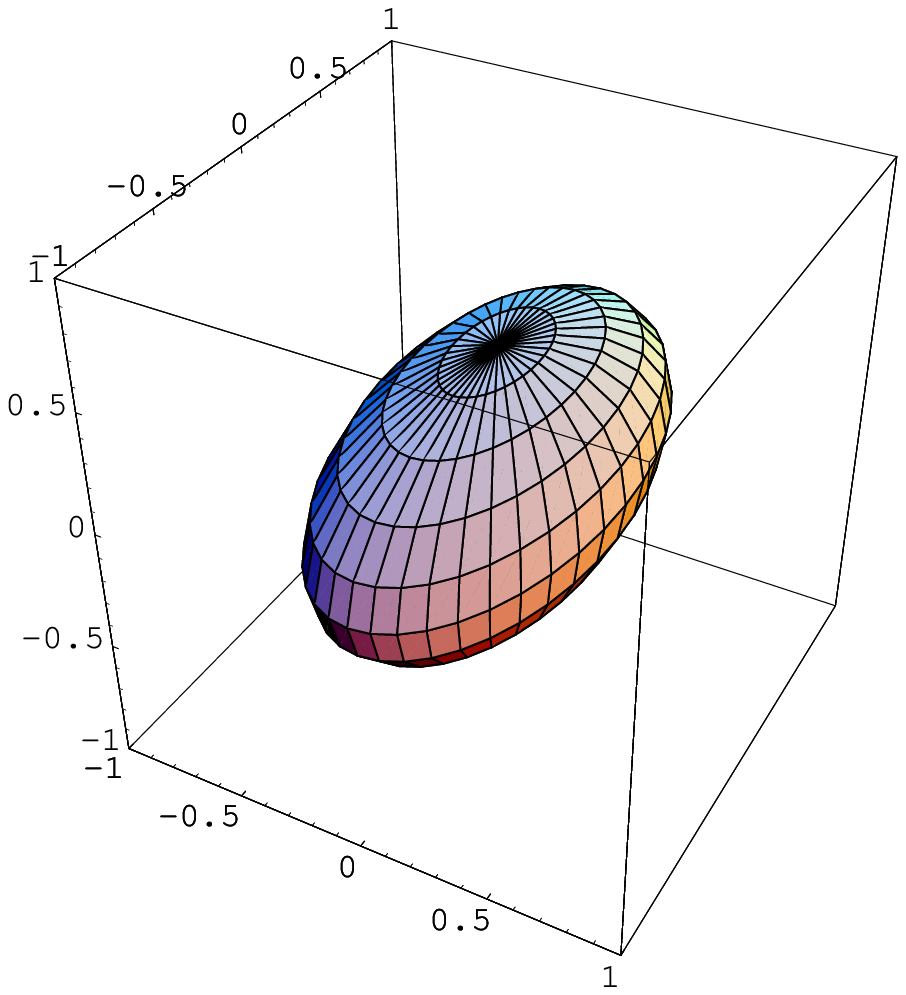}
\\
\includegraphics[width=2.4cm]{SferaBloch.eps}
\includegraphics[width=2.4cm]{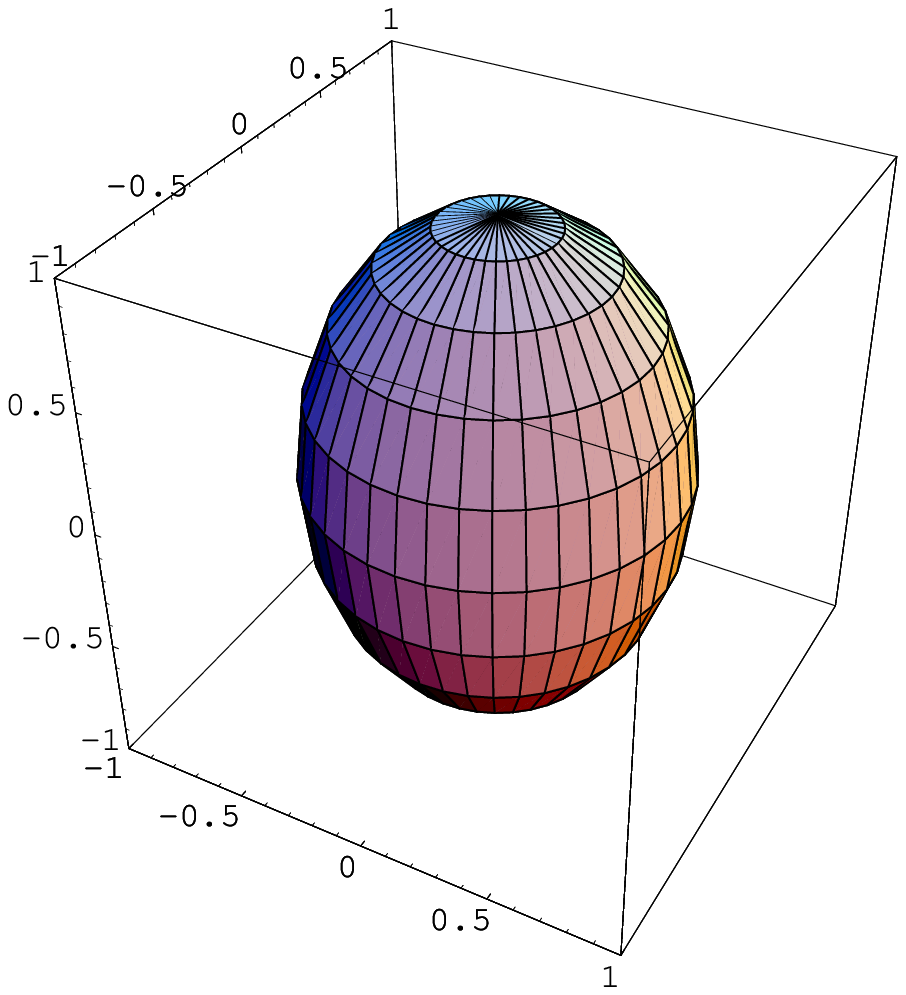}
\includegraphics[width=2.4cm]{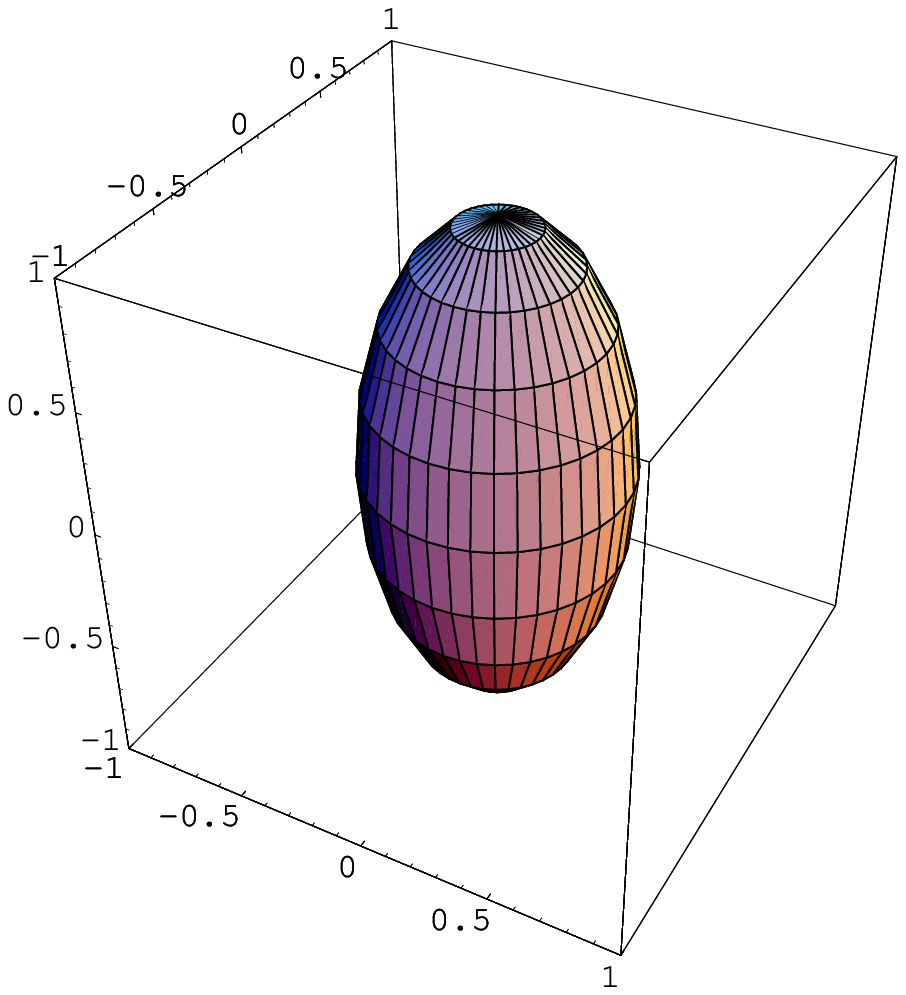}
\end{center}
\caption{Graphic visualization of deformations of the Bloch sphere along the $x$, $y$ and $z$ axes by means of the bit flip channel, the bit-phase flip channel and the phase flip channel, respectively. The Bloch sphere (left) represents initial unperturbed states, while the ellipsoids (right) show how the Bloch sphere is affected by the noise channels, for increasing values of the error parameter ($\theta = 0, \frac{\pi}{4}, \frac{\pi}{3}$).}\label{app-gv-def}
\end{figure}

These transformations can be finally visualized in the Bloch sphere representation (Figure \ref{app-gv-def}).

\subsection{Displacements of the Bloch sphere along the coordinate axes}

As happens for deformations, also displacement errors are non-unitary errors and, consequently, require a physically feasible representation involving a unitary operation in an extended quantum system.
The most important feature of all displacement channels is that they must perform at the same time a correlated deformation, so that the resulting matrix $\rho^\prime$ is still a density matrix representing a physically feasible state.

\begin{figure}[!htb]
\begin{center}
\includegraphics[width=10.8cm]{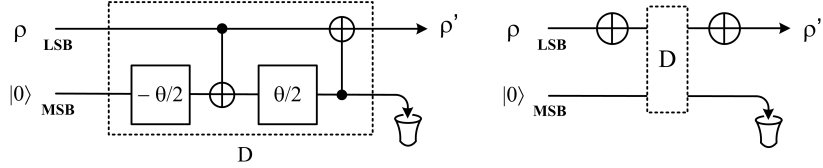}
\end{center}
\caption{Quantum circuits representing displacements of the Bloch sphere along the axis $z$. From left to right, the amplitude damping channel and its reversion to achieve the displacement in the positive and negative directions of the axis $z$, respectively.}\label{sd-II-qcampldamp}
\end{figure}

\begin{figure}[!htb]
\begin{center}
\includegraphics[width=9.4cm]{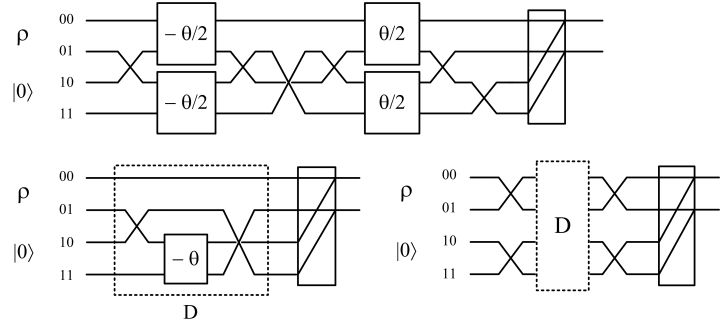}
\end{center}
\caption{Diagrams of states representing displacements of the Bloch sphere along the axis $z$. From top to bottom, from left to right: complete and simplified diagrams of states of the amplitude damping channel, and diagram of states of its reversion.}\label{sd-II-sdampldamp}
\end{figure}

\begin{figure}[!htb]
\begin{center}
\includegraphics[width=10.6cm]{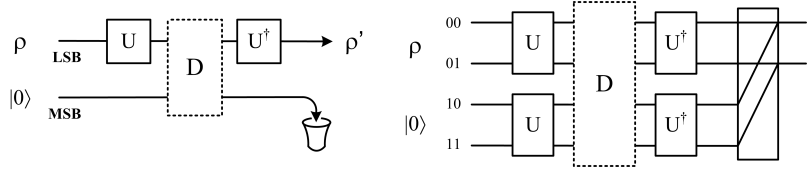}
\end{center}
\caption{Quantum circuits (left) and diagrams of states (right) representing displacements of the Bloch sphere along the axes $x$ and $y$. Circuits and diagrams are derived by rotating the displacement direction along which the amplitude damping channel and its reversion act. Rotations are obtained by applying the appropriate unitary matrices $U$ and $U^{\dagger}$, while \qq D'' denotes the gate array corresponding to the amplitude damping channel, as shown in Figures \ref{sd-II-qcampldamp} and \ref{sd-II-sdampldamp}.}\label{sd-II-displxy}
\end{figure}

The transformations induced by displacements of the center of the Bloch sphere along the axes $x, y$ or $z$ can be implemented by means of the amplitude damping channel and its appropriate modifications to reverse or rotate the displacement direction\footnote{We remark that the amplitude damping channel actually describes the most general deformation errors: It performs the best possible deformation of the Bloch sphere in relation to the desired displacement of its center along the considered coordinate axis, as given by the high degree of tangency of the resulting ellipsoid to the initial Bloch sphere.}.
The quantum circuits for the standard and modified amplitude damping channels are illustrated in Figures \ref{sd-II-qcampldamp} and \ref{sd-II-displxy}(left), where the environment is again represented by a single ancillary qubit. The processing of information caused by the different instances of the amplitude damping channel is clearly illustrated by the diagrams of states in Figures \ref{sd-II-sdampldamp} and \ref{sd-II-displxy} (right).

The Kraus operators for each instance of the amplitude damping channel can be easily derived from the corresponding diagram of states.
For the standard amplitude damping channel we have:
\begin{equation}
F_0=
\left[
\begin{array}{cc}
1 & 0\\
0 & \cos\frac{\theta}{2}
\end{array}
\right], \mh F_1=
\left[
\begin{array}{cc}
0 & \sin\frac{\theta}{2} \\
0 & 0
\end{array}
\right].
\end{equation}
For the reversed amplitude damping channel, we have:
\begin{equation}
F_0=
\left[
\begin{array}{cc}
\cos\frac{\theta}{2} & 0\\
0 & 1
\end{array}
\right], \mh F_1=
\left[
\begin{array}{cc}
0 & 0 \\
\sin\frac{\theta}{2} & 0
\end{array}
\right].
\end{equation}
To rotate the amplitude damping channel for displacements along the axes $x$ and $y$, we apply respectively the unitary transformations:
\begin{equation}
    U_{(x)}=\frac{1}{\sqrt{2}}
\left[
\begin{array}{cc}
1 & \pm 1\\
\mp 1 & 1
\end{array}
\right], \mh U_{(y)}=\frac{1}{\sqrt{2}}
\left[
\begin{array}{cc}
1 & \pm i\\
\pm i & 1
\end{array}
\right].
\end{equation}
Thus we obtain the following Kraus operators for $x$ displacements:
$$
F_0=
\frac{1}{2}\left[
\begin{array}{cc}
1+\cos\frac{\theta}{2} & \pm (1-\cos\frac{\theta}{2})\\
\pm (1-\cos\frac{\theta}{2}) & 1+\cos\frac{\theta}{2}
\end{array}
\right],
$$
\begin{equation}
F_1=
\frac{1}{2}\left[
\begin{array}{cc}
\mp \sin\frac{\theta}{2} & \sin\frac{\theta}{2} \\
-\sin\frac{\theta}{2} & \pm \sin\frac{\theta}{2}
\end{array}
\right]
\end{equation}
and for $y$ displacements:
$$ F_0=
\frac{1}{2}\left[
\begin{array}{cc}
1+\cos\frac{\theta}{2} & \pm i (1-\cos\frac{\theta}{2})\\
\mp i (1-\cos\frac{\theta}{2}) & 1+\cos\frac{\theta}{2}
\end{array}
\right],
$$
\begin{equation}
F_1=
\frac{1}{2}\left[
\begin{array}{cc}
\pm i \sin\frac{\theta}{2} & \sin\frac{\theta}{2} \\
\sin\frac{\theta}{2} & \mp i \sin\frac{\theta}{2}
\end{array}
\right].
\end{equation}

The Kraus operators induce the following transformations in the Bloch sphere coordinates of the initial state, respectively for the negative and positive directions of the axes $x, y, z$:
$$
    \left\{
\begin{array}{l}
X^\prime=\mp \sin^2 \theta \pm \cos^2 \theta X\\
Y^\prime=\cos \theta Y\\
Z^\prime=\cos \theta Z
\end{array}
\right.
\mh
\left\{
\begin{array}{l}
X^\prime=\cos \theta X\\
Y^\prime=\mp \sin^2 \theta \pm \cos^2 \theta Y\\
Z^\prime=\cos \theta Z
\end{array}
\right.
$$
\begin{equation}
    \left\{
\begin{array}{l}
X^\prime=\cos \theta X\\
Y^\prime=\cos \theta Y\\
Z^\prime=\mp \sin^2 \theta \pm \cos^2 \theta Z
\end{array}
\right.
\end{equation}

\begin{figure}[!p]
\begin{center}
\includegraphics[width=2.4cm]{SferaBloch.eps}
\includegraphics[width=2.4cm]{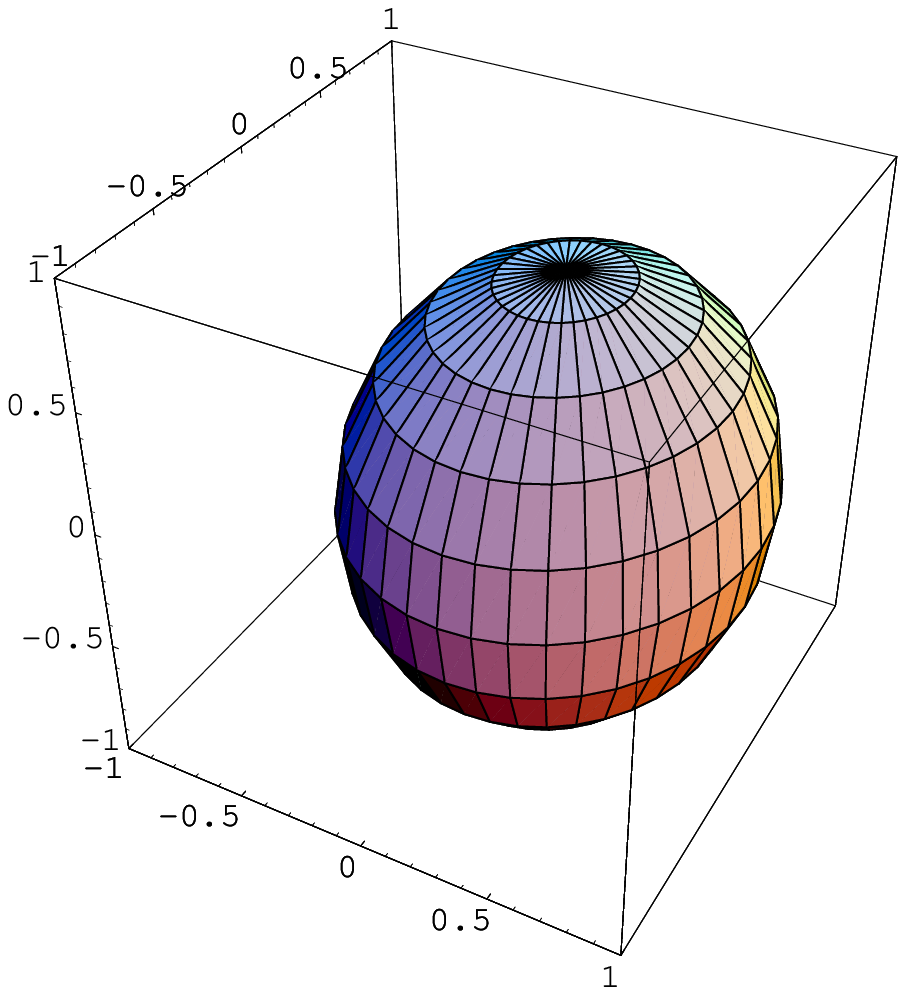}
\includegraphics[width=2.4cm]{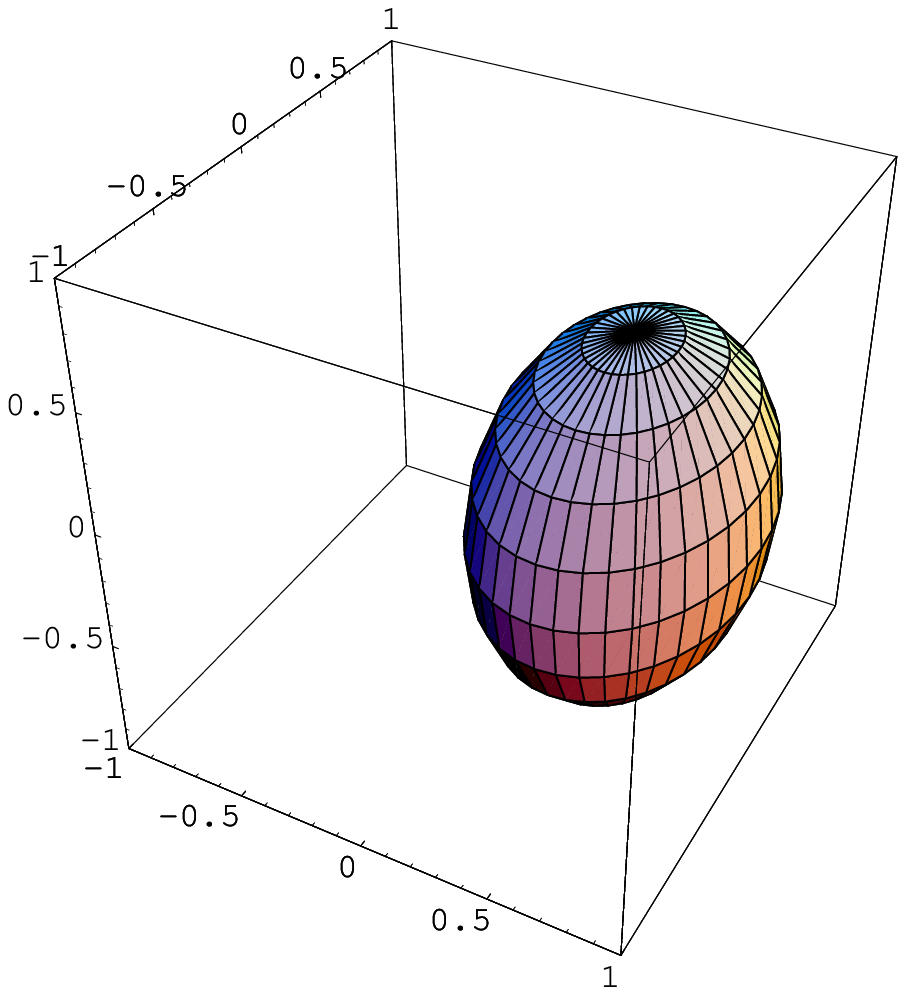}
\includegraphics[width=2.4cm]{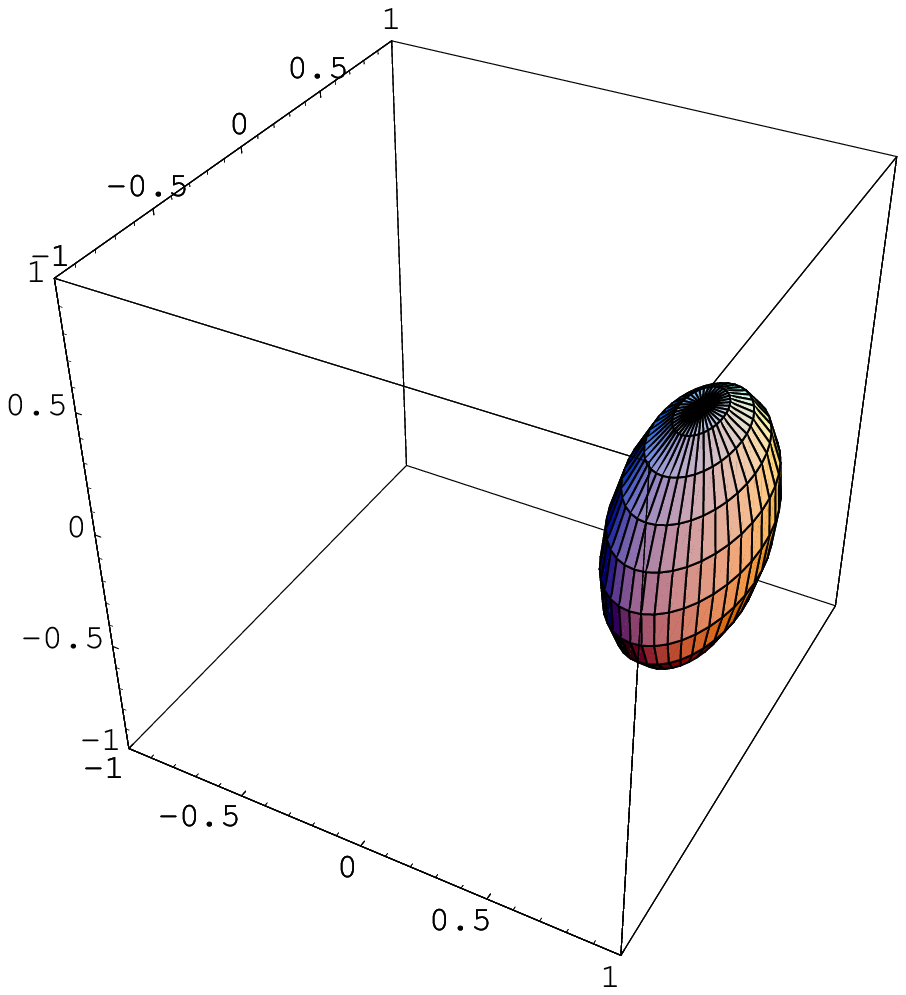}
\\
\includegraphics[width=2.4cm]{SferaBloch.eps}
\includegraphics[width=2.4cm]{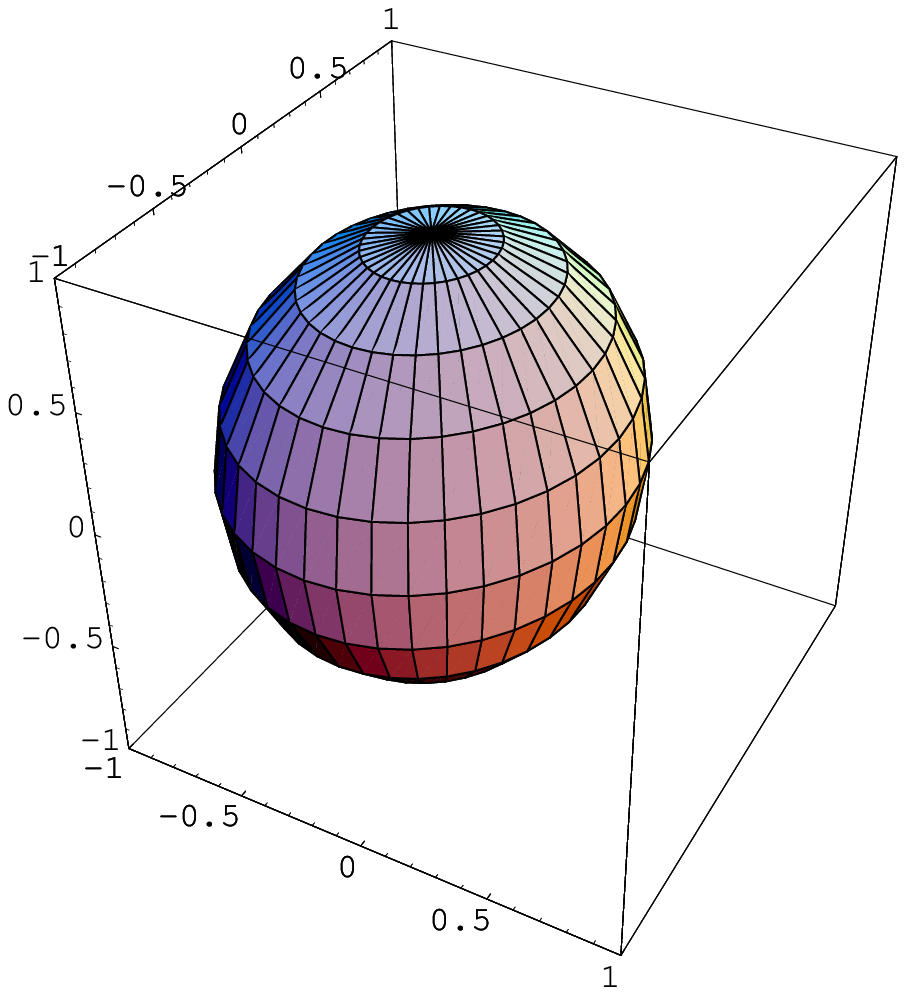}
\includegraphics[width=2.4cm]{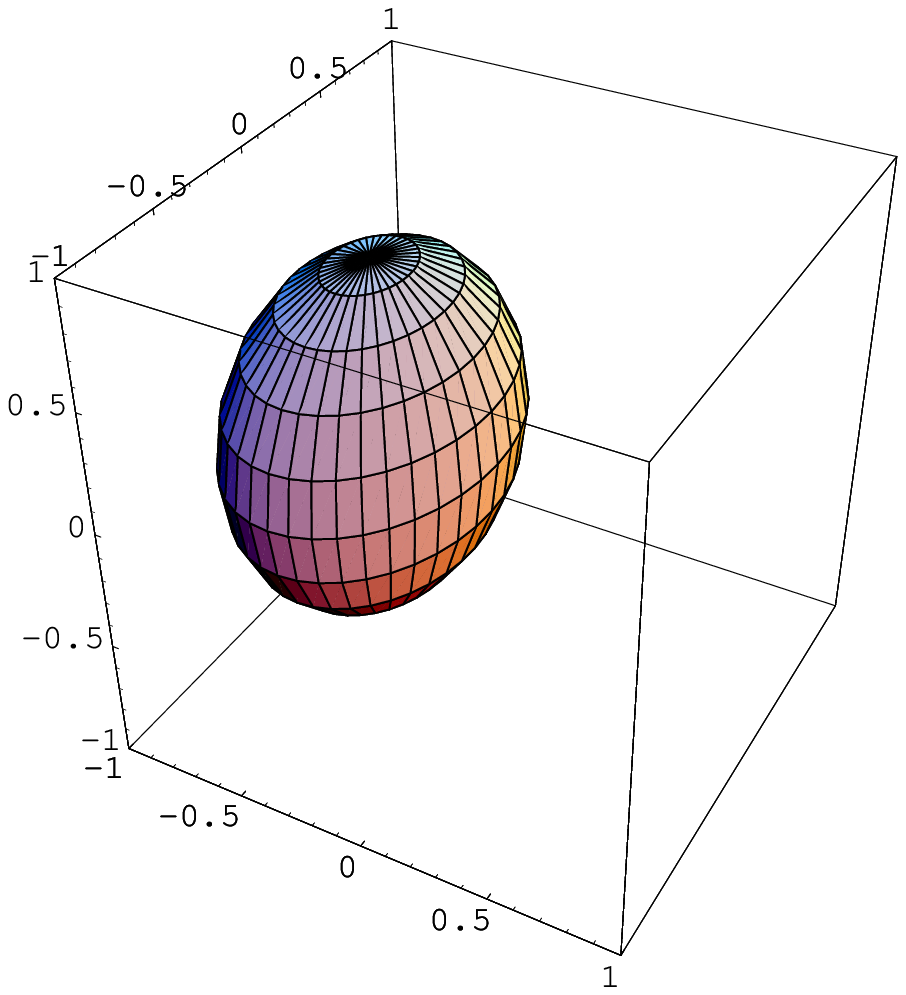}
\includegraphics[width=2.4cm]{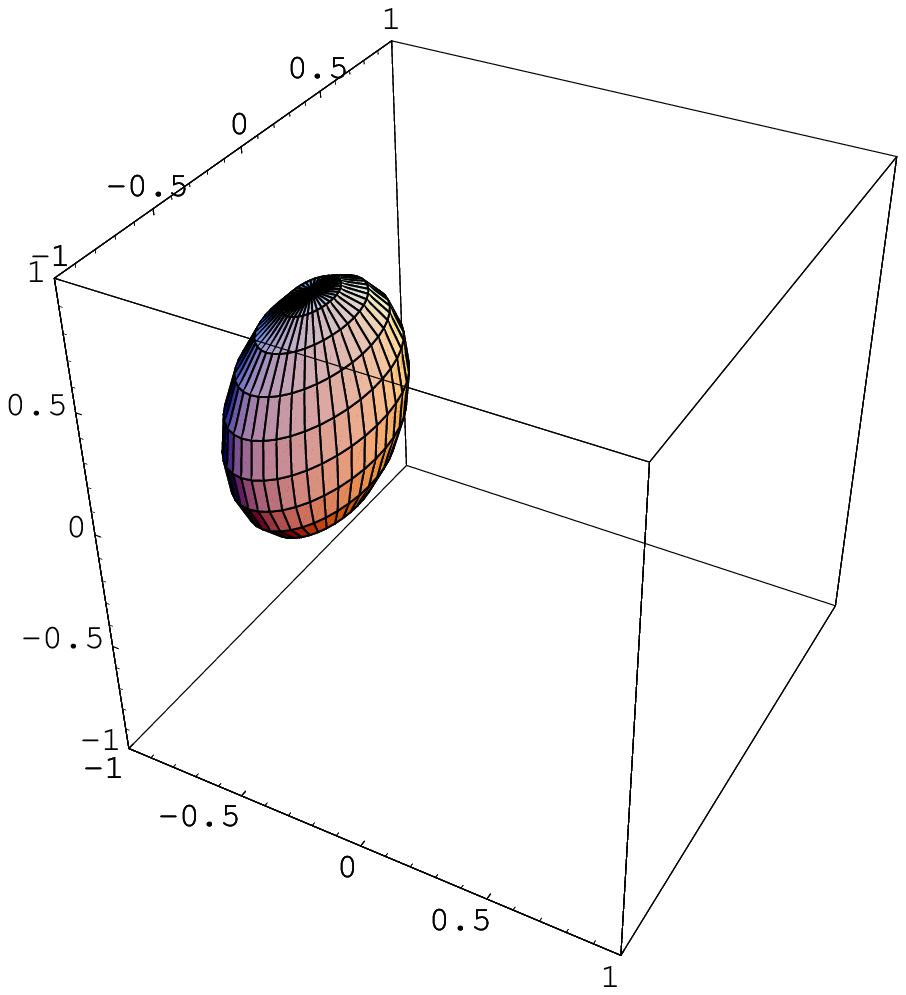}
\\
\includegraphics[width=2.4cm]{SferaBloch.eps}
\includegraphics[width=2.4cm]{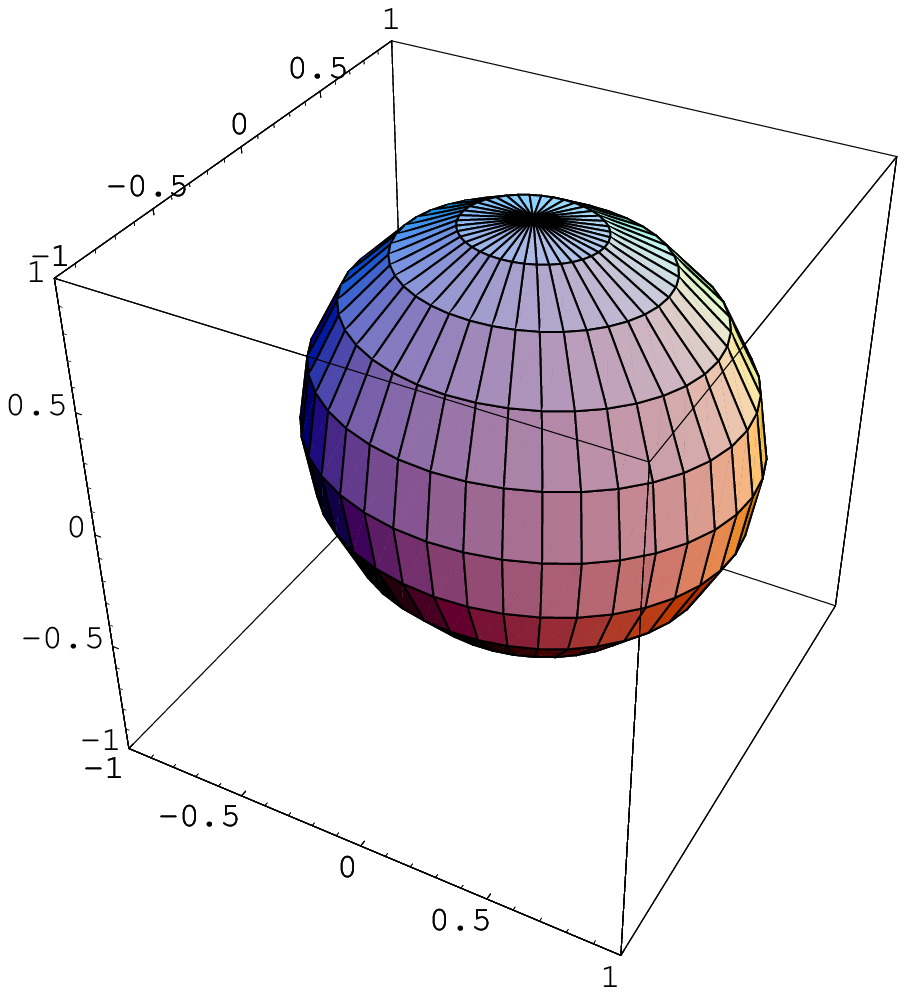}
\includegraphics[width=2.4cm]{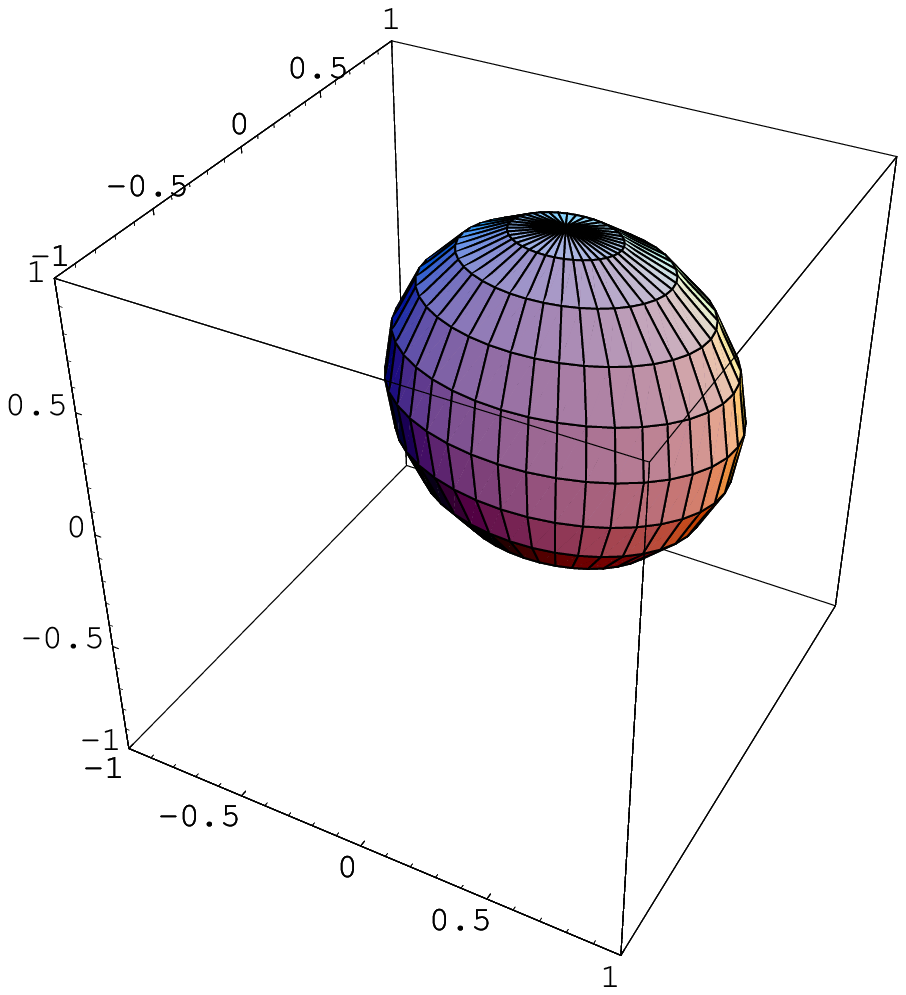}
\includegraphics[width=2.4cm]{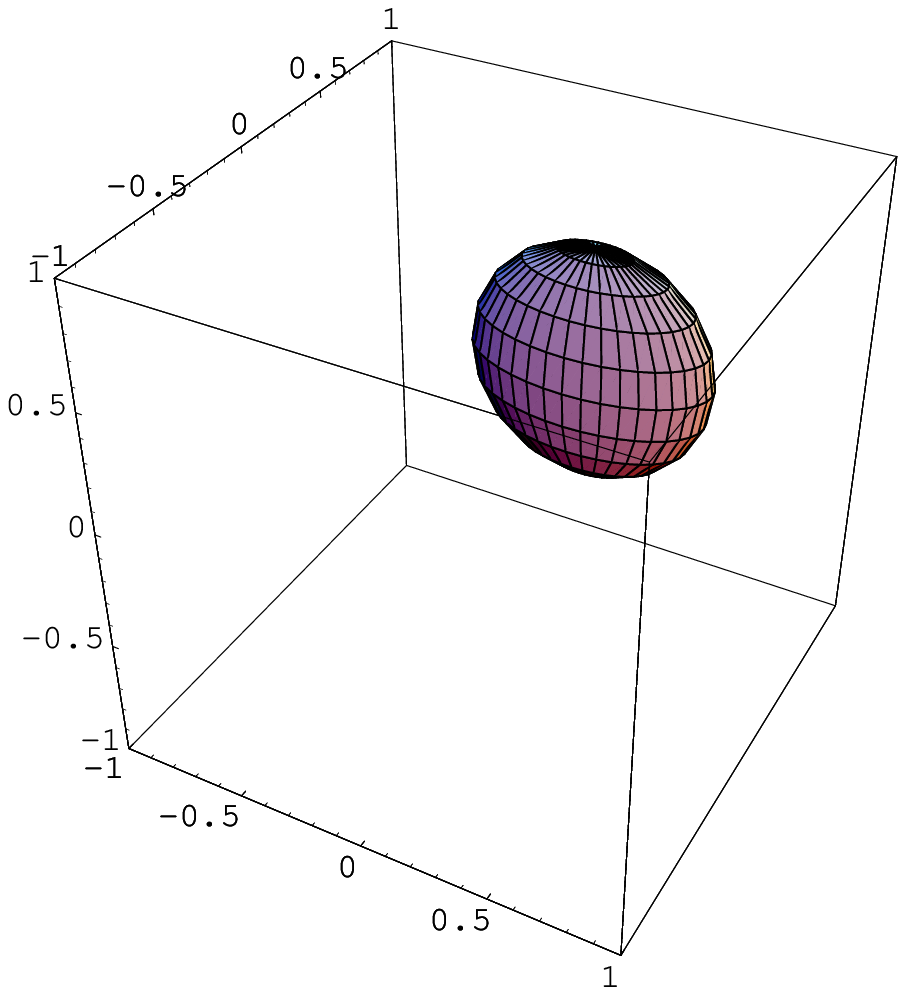}
\\
\includegraphics[width=2.4cm]{SferaBloch.eps}
\includegraphics[width=2.4cm]{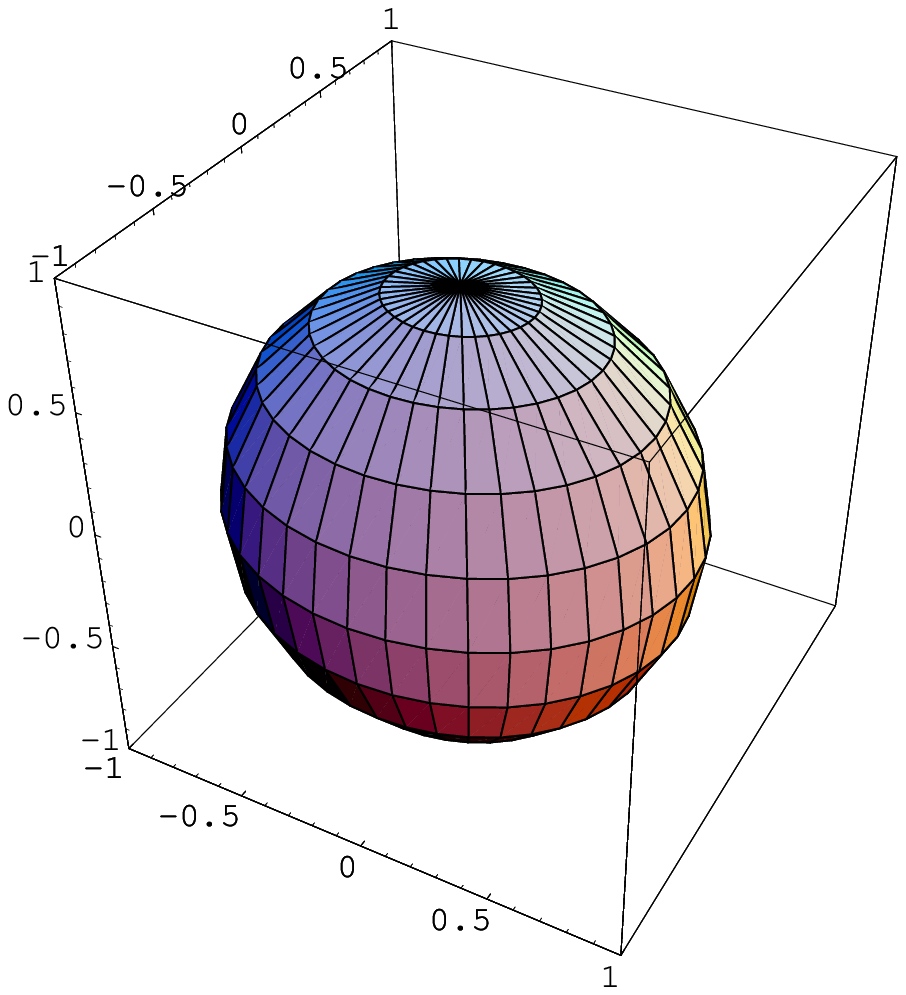}
\includegraphics[width=2.4cm]{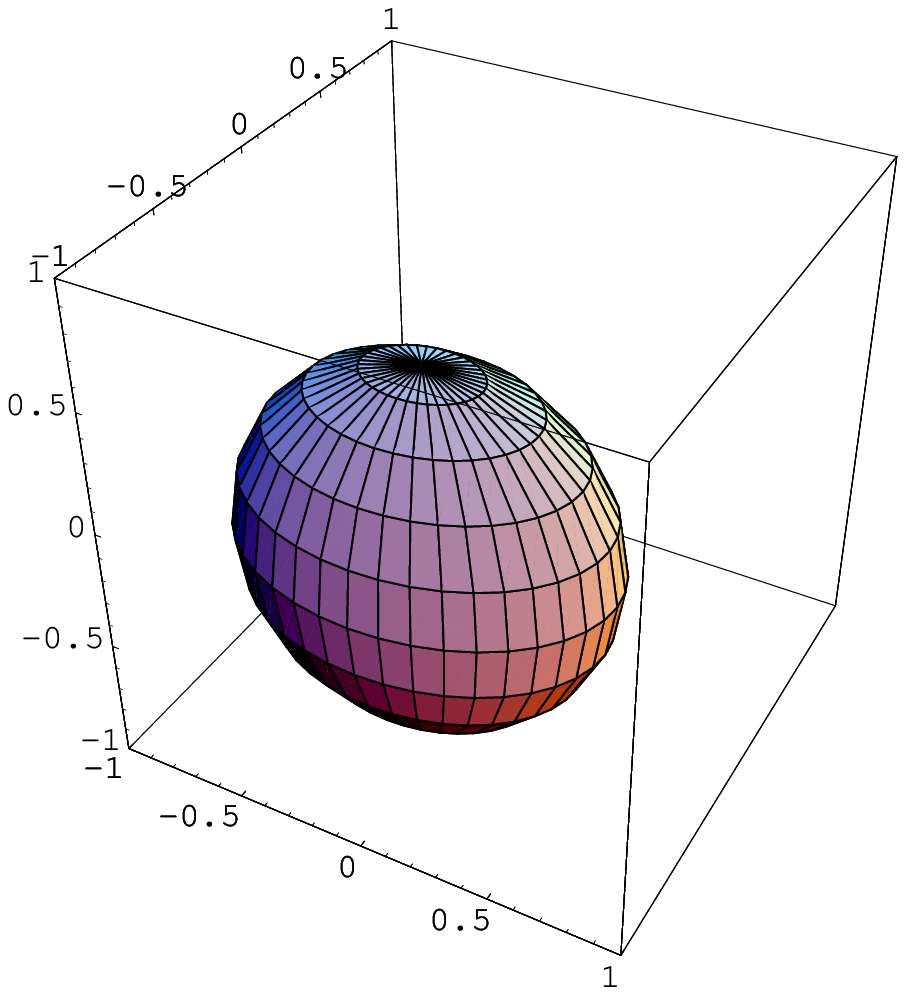}
\includegraphics[width=2.4cm]{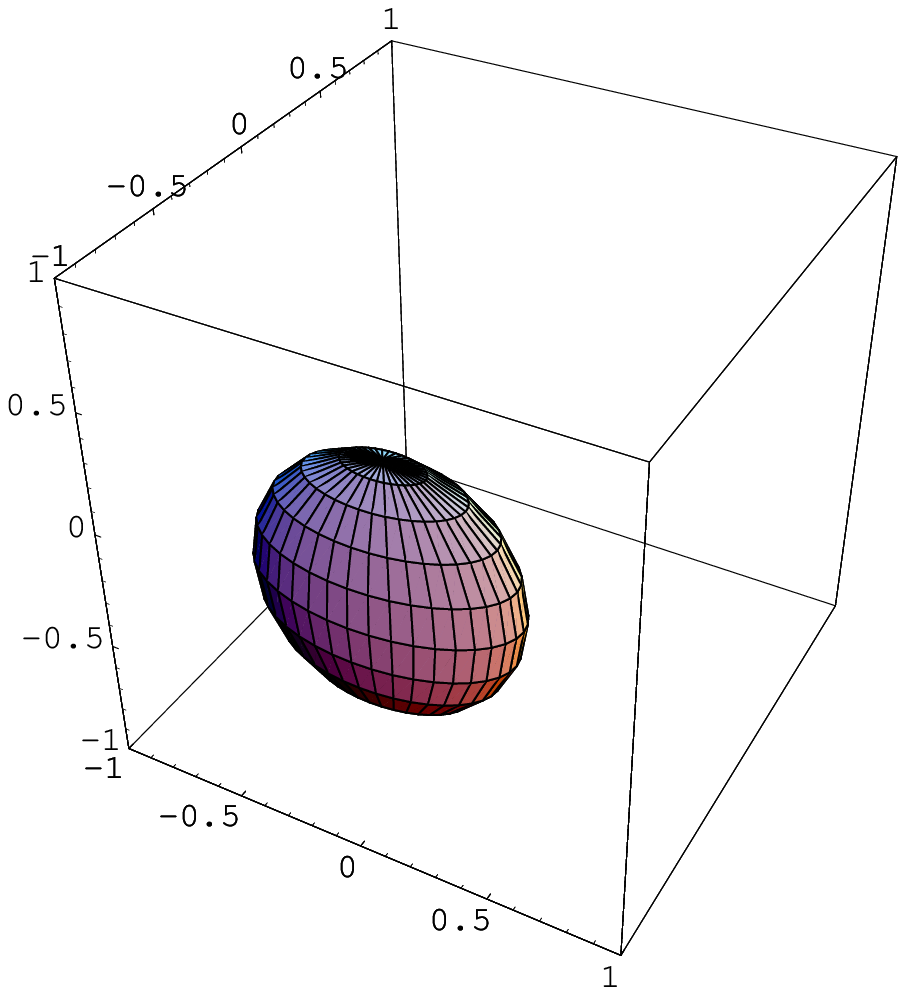}
\\
\includegraphics[width=2.4cm]{SferaBloch.eps}
\includegraphics[width=2.4cm]{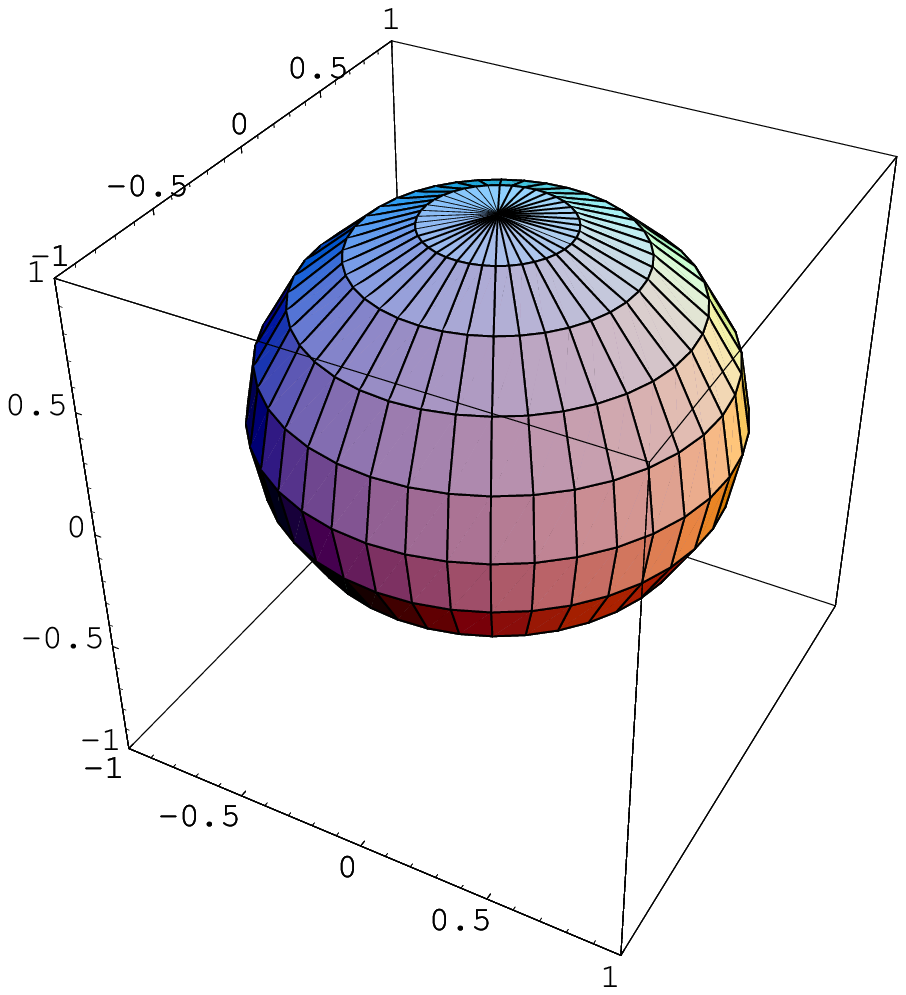}
\includegraphics[width=2.4cm]{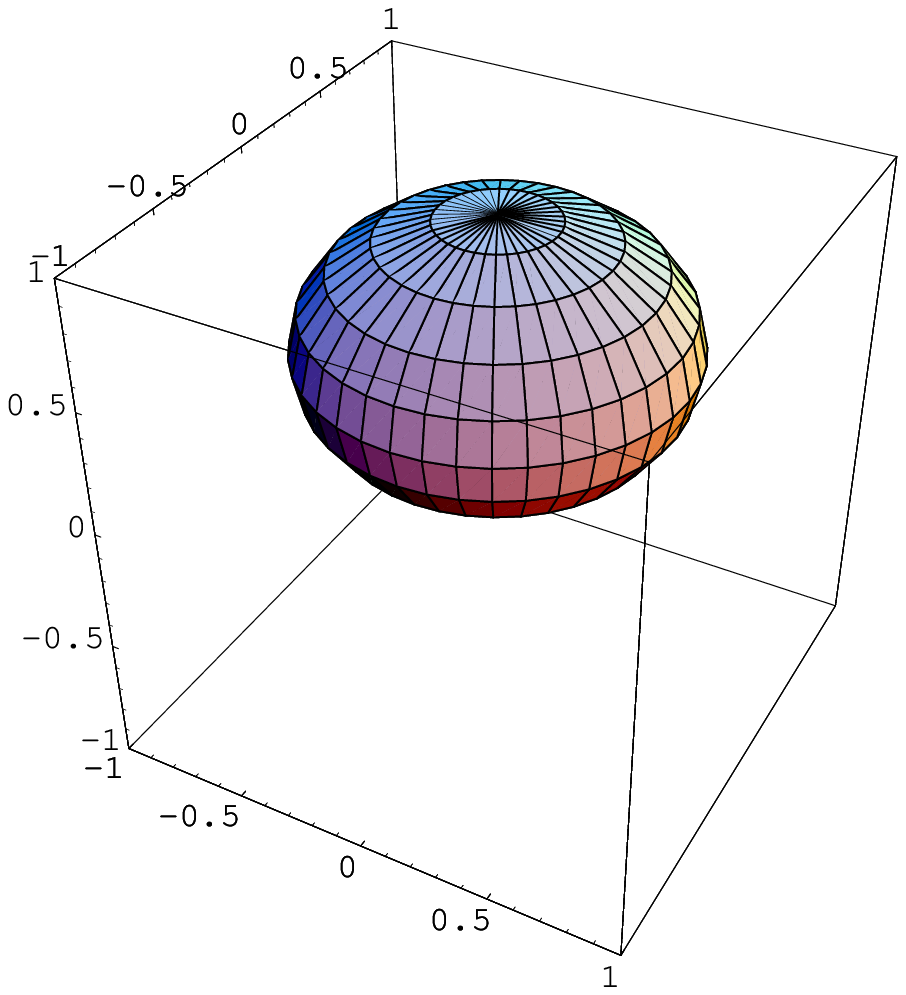}
\includegraphics[width=2.4cm]{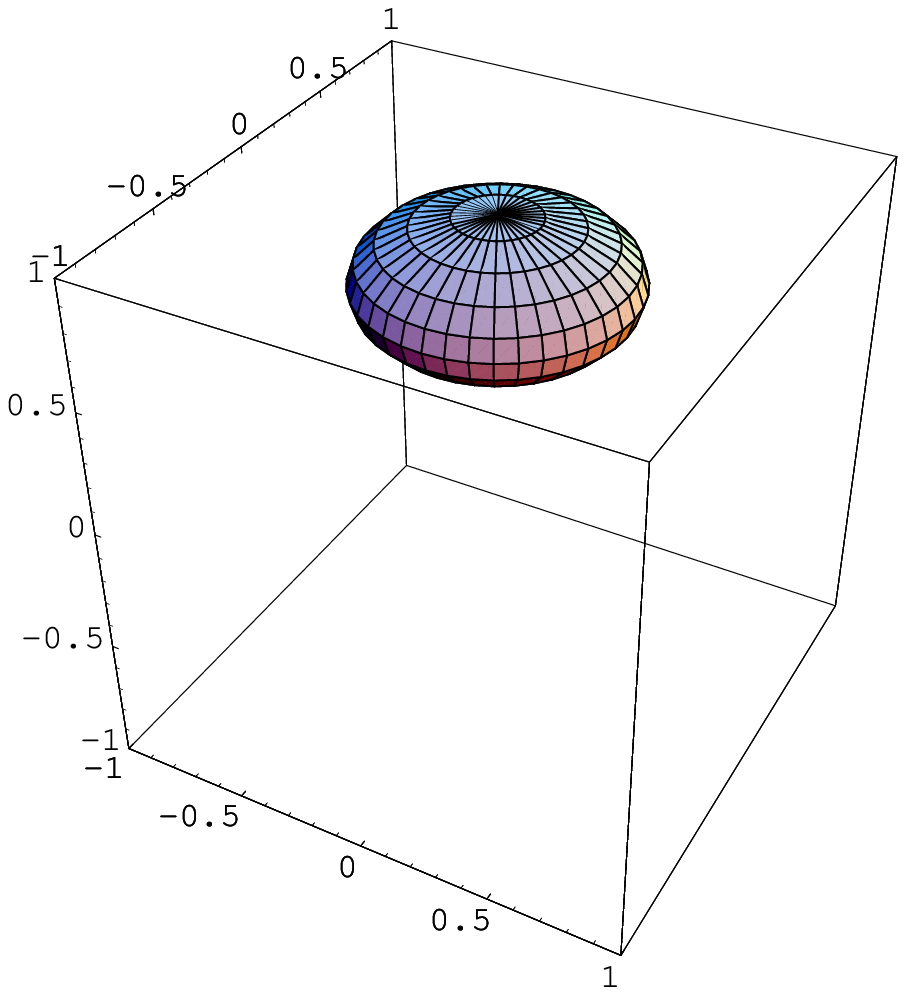}
\\
\includegraphics[width=2.4cm]{SferaBloch.eps}
\includegraphics[width=2.4cm]{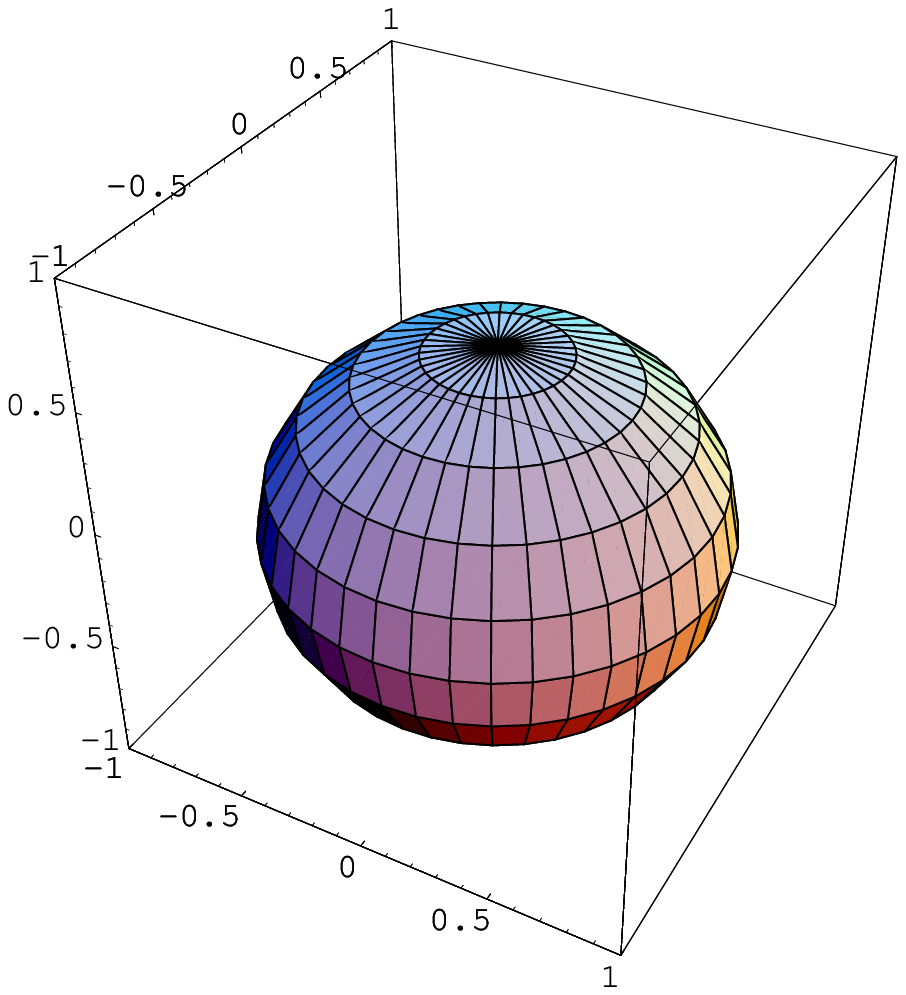}
\includegraphics[width=2.4cm]{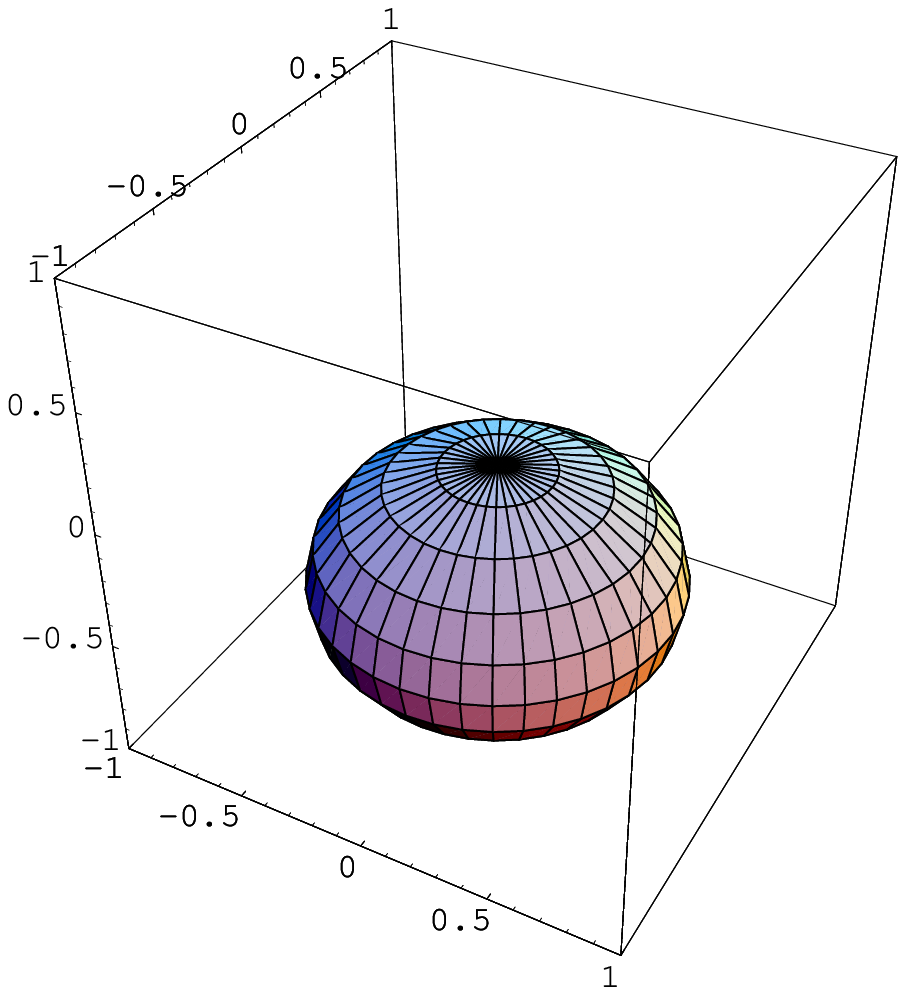}
\includegraphics[width=2.4cm]{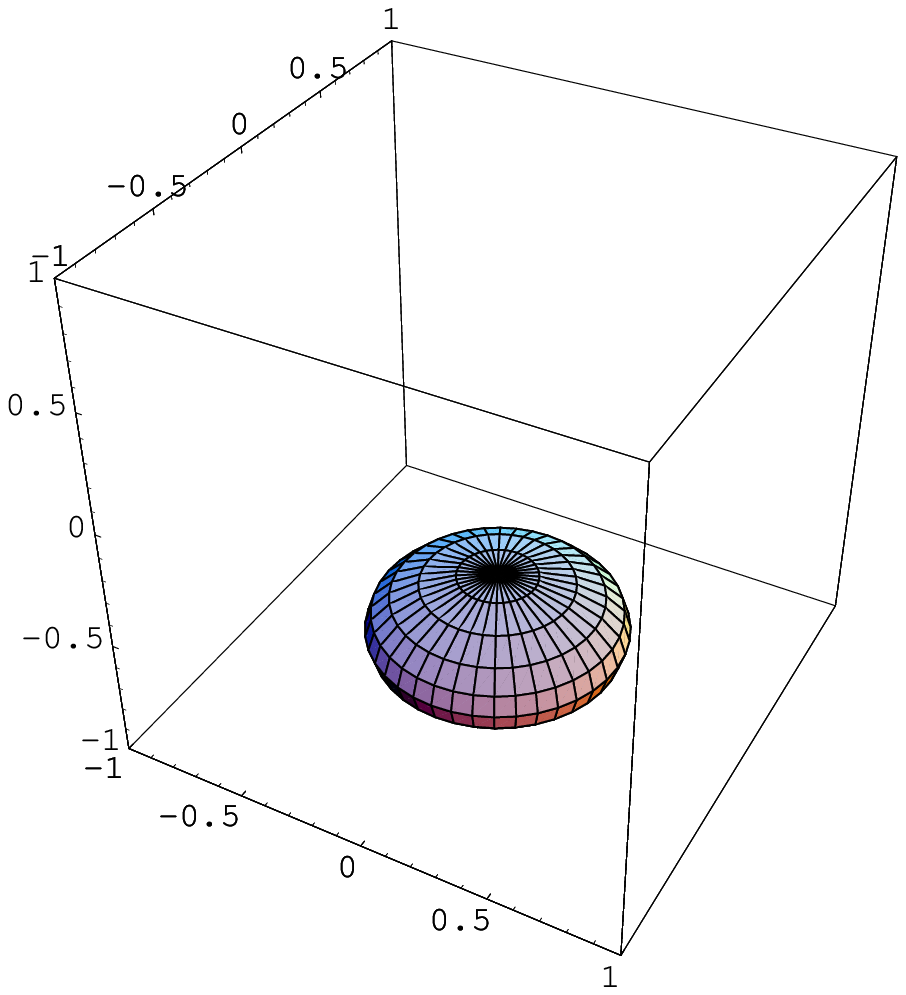}
\end{center}
\caption{Graphic visualization of displacements of the Bloch sphere along the $x$, $y$ and $z$ axis. The Bloch sphere (left) represents initial unperturbed states, while the ellipsoids (right) show how the Bloch sphere is affected by the noise channel, for increasing values of the error parameter ($\theta = 0, \frac{\pi}{6},
\frac{\pi}{4}, \frac{\pi}{3}$).}\label{app-gv-displ}
\end{figure}

These transformations can be finally visualized in the Bloch sphere representation in Figure \ref{app-gv-displ}.

\subsection{The depolarizing channel}

\begin{figure}[!htb]
\begin{center}
\includegraphics[width=10.4cm]{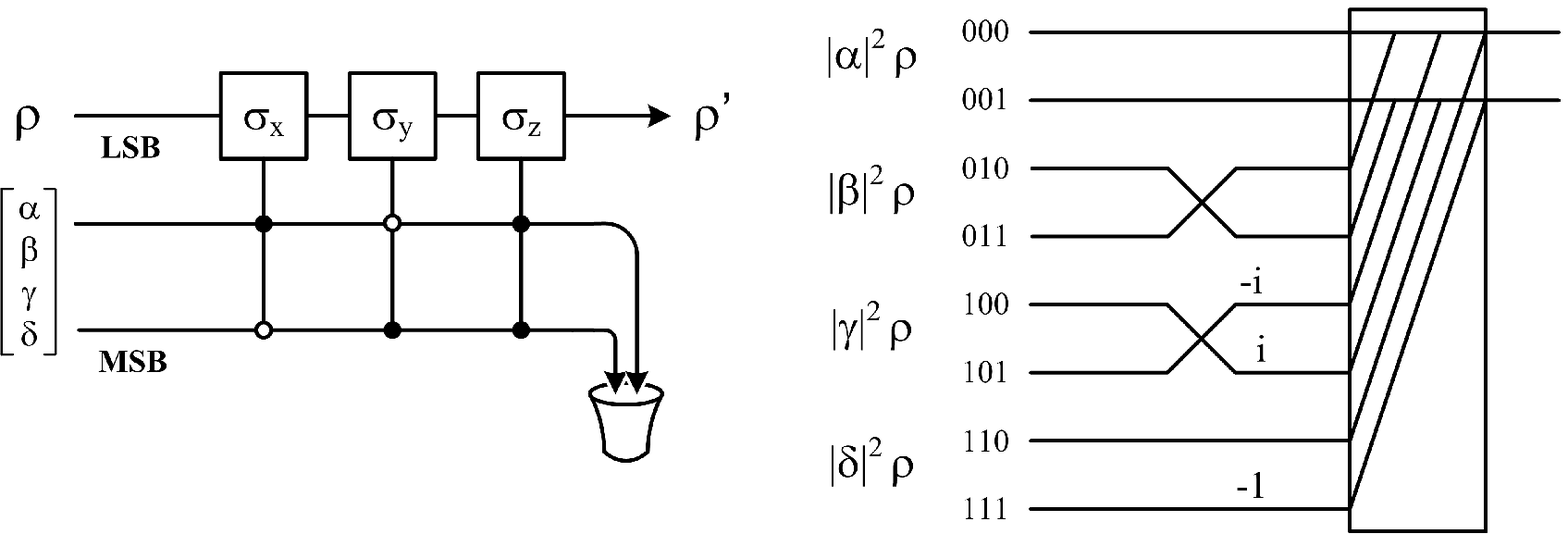}
\end{center}
\caption{Quantum circuit (left) and diagram of states (right) representing a generalized depolarizing channel. Depending on the choice of the parameters $|\alpha|, |\beta|$ and $|\gamma|$, several noise operations can be described as special cases of this channel.}\label{sd-II-depolch}
\end{figure}

After describing the complete set of basic single-qubit noise operations, we conclude this section by mentioning a well known and widely used single-qubit decoherence model: the depolarizing channel (see, \textit{e.g.}, \cite{qcbook2}, pages 340-343).
The depolarizing channel can be described by means of the quantum circuit in Figure \ref{sd-II-depolch} (left).

The processing of information performed by the depolarizing channel is clearly illustrated in the diagram of states in Figure \ref{sd-II-depolch} (right).
The main single-qubit system is coupled with the environment, which is now represented by two ancillary qubits, set in the pure state:
\begin{equation}
\ket \Psi = \alpha |00\rangle + \beta |01\rangle + \gamma |10\rangle + \delta |11\rangle,
\end{equation}
with the normalization condition:
\begin{equation}
|\alpha|^2 + |\beta|^2 + |\gamma|^2 + |\delta|^2 = 1.
\end{equation}
The channel applies one of the the operators $I, \sigma_x, \sigma_y$ or $\sigma_z$ to the least significant qubit if the two most significant qubits are in one (or in a superposition) of the states $|00\rangle, |01\rangle, |10\rangle$ or $|11\rangle$.
The final state of the single-qubit system is obtained by tracing over environmental qubits.

Depending on the choice of the parameters $|\alpha|, |\beta|$ and $|\gamma|$, several noise operations can be described as special cases of this channel (for instance, the bit flip or the phase flip channels).

The standard depolarizing channel is obtained when:
\begin{equation}
|\alpha|^2=\cos^2\theta, \sh |\beta|^2 =
|\gamma|^2 = |\delta|^2 = \frac{\sin^2\theta}{3}.
\end{equation}
In this case, the Kraus operators become:
$$
    F^{(1)}= \cos \theta \, I, \mh
    F^{(2)}=\frac{\sin\theta}{\sqrt3}
 \, \sigma_x,
$$
\begin{equation}
F^{(3)}=\frac{\sin\theta}{\sqrt3}  \, \sigma_y, \mh
F^{(4)}=\frac{\sin\theta}{\sqrt3}  \, \sigma_z,
\end{equation}
and the Bloch sphere coordinates of the initial state are transformed as follows:
\begin{equation}
 \left\{
\begin{array}{l}
X^\prime= \left(1-\frac{4}{3}\sin^2\theta\right) X\\
Y^\prime= \left(1-\frac{4}{3}\sin^2\theta\right) Y\\
Z^\prime= \left(1-\frac{4}{3}\sin^2\theta\right) Z
\end{array}
\right.
\end{equation}
Thus, the Bloch sphere is deformed into an ellipsoid,
centered at the origin of the Bloch sphere, whose axes are directed along
$x, y$ and $z$, the deformation rate being the same along each axis.

Thus, the depolarizing channel can not be considered as a general model to describe and study single-qubit decoherence effects, since it only includes a small subset of specific noise operations, although involving an higher dimensional auxiliary system in respect to the previously illustrated single-qubit noise channels.


\section{Conclusions and Future Developments}\label{sec-sd-II-concl}

We have explored the main processes involved in evolution of general quantum systems by means of \textit{Diagrams of States}, a novel method to graphically represent and analyze how quantum information is elaborated during computations performed by quantum circuits.
We have offered complete and detailed descriptions by diagrams of states of partial trace operations, density matrix purification and evolution by Kraus operators, single-qubit decoherence and noise channels.

In our opinion, diagrams of states can be used as an auxiliary or as an alternative approach to standard methods, both to investigate and to conceive quantum computations. Analytical study and Feynman diagrams alone are often too synthetic to clearly visualize how quantum information is processed by computations. On the contrary, the dimension of the graphic representation of states grows exponentially in respect to the dimension of the examined quantum system, thus offering a complete and detailed visualization of the computational process.

Diagrams of states appear to be most useful whenever the quantum operations to be analyzed are described by very sparse matrices, since only non-null entries of matrices are associated with diagram lines which contain active information. This way, the resulting diagrams show clearly and immediately the significant pattern along which quantum information is processed by the computation from input to output. Indeed, several quantum computations actually involve operations satisfying this requirement, and evidence of this is also provided by the processes illustrated in this paper.

Further computations are going to be explored by this graphic representation, among which quantum algorithms \cite{FeStsdIV} and entanglement-based practical applications \cite{FeStsdIII}.

\section*{Acknowledgments}

The authors wish to thank Samuel L. Braunstein
and Roberto Suardi for their kind contributions to the development and improvement of this paper.

Sara Felloni acknowledges support by ERCIM, as this work was partially carried out during the tenure of an ERCIM \qq Alain Bensoussan'' Fellowship Programme.

\end{document}